\documentclass[twocolumn]{aastex631}

\usepackage{amsmath}

\begin{document}

\title{Detection of a Spatially Extended Stellar Population in M33: A Shallow Stellar Halo?}

\correspondingauthor{Itsuki Ogami}
\email{itsuki.ogami@grad.nao.ac.jp}

\author[0000-0001-8239-4549]{Itsuki Ogami}
\affiliation{The Graduate University for Advanced Studies (SOKENDAI), 2-21-1 Osawa, Mitaka, Tokyo 181-8588, Japan}
\affiliation{National Astronomical Observatory of Japan, 2-21-1 Osawa, Mitaka, Tokyo 181-8588, Japan}

\author[0000-0002-3852-6329]{Yutaka Komiyama}
\affiliation{Department of Advanced Sciences, Faculty of Science and Engineering, Hosei University, 3-7-2 Kajino-cho, Koganei, Tokyo 184-8584, Japan}

\author[0000-0002-9053-860X]{Masashi Chiba}
\affiliation{Astronomical Institute, Tohoku University, Aoba-ku, Sendai, Miyagi 980-8578, Japan}

\author[0000-0003-2258-7044]{Mikito Tanaka}
\affiliation{Department of Advanced Sciences, Faculty of Science and Engineering, Hosei University, 3-7-2 Kajino-cho, Koganei, Tokyo 184-8584, Japan}

\author[0000-0001-8867-4234]{Puragra Guhathakurta}
\affiliation{Department of Astronomy and Astrophysics, University of California Santa Cruz, University of California Observatories, 1156 High Street, Santa Cruz, CA 95064, USA}

\author[0000-0001-6196-5162]{Evan N. Kirby}
\affiliation{Department of Physics and Astronomy, University of Notre Dame, Notre Dame, IN 46556, USA}

\author[0000-0002-4013-1799]{Rosemary F.G. Wyse}
\affiliation{Department of Physics and Astronomy, Johns Hopkins University, Baltimore, MD 21218, USA}

\author[0000-0001-5522-5029]{Carrie Filion}
\affiliation{Department of Physics and Astronomy, Johns Hopkins University, Baltimore, MD 21218, USA}

\author[0000-0001-6503-8315]{Takanobu Kirihara}
\affiliation{Kitami Institute of Technology, 165, Koen-cho, Kitami, Hokkaido 090-8507, Japan}

\author[0000-0003-4656-0241]{Miho N. Ishigaki}
\affiliation{National Astronomical Observatory of Japan, 2-21-1 Osawa, Mitaka, Tokyo 181-8588, Japan}

\author[0000-0002-8758-8139]{Kohei Hayashi}
\affiliation{National Institute of Technology, Sendai College, Natori, Miyagi 981-1239, Japan}
\affiliation{Astronomical Institute, Tohoku University, Aoba-ku, Sendai, Miyagi 980-8578, Japan}
\affiliation{Institute for Cosmic Ray Research, The University of Tokyo, Kashiwa, Chiba 277-8582, Japan}

\begin{abstract}
We analyze the outer regions of M33, beyond 15 kpc in projected distance from its center using Subaru/HSC multi-color imaging. We identify Red Giant Branch (RGB) stars and Red Clump (RC) stars using the surface gravity sensitive {\it NB515} filter for the RGB sample, and a multi-color selection for both samples. We construct the radial surface density profile of these RGB and RC stars, and find that M33 has an extended stellar population with a shallow power-law index of $\alpha > -3$, depending on the intensity of the contamination. This result represents a flatter profile than the stellar halo which has been detected by the previous study focusing on the central region, suggesting that M33 may have a double-structured halo component, i.e. inner/outer halos or a very extended disk. Also, the slope of this extended component is shallower than those typically found for halos in large galaxies, implying intermediate-mass galaxies may have different formation mechanisms (e.g., tidal interaction) from large spirals. We also analyze the radial color profile of RC/RGB stars, and detect a radial gradient, consistent with the presence of an old and/or metal-poor population in the outer region of M33, thereby supporting our proposal that the stellar halo extends beyond 15 kpc. Finally, we estimate that the surface brightness of this extended component is $\mu_{\it V} = 35.72 \pm 0.08$ mag arcsec$^{-2}$. If our detected component is the stellar halo, this estimated value is consistent with the detection limit of previous observations.
\end{abstract}


\section{Introduction} \label{section:intro}
In the $\Lambda$-dominated cold dark matter scenario, mergers play an important role in galaxy formation. The way in which galaxies build their mass depends mostly on this process \citep{bullock2005}. After an accretion event, accreted stars are distributed in the outer region of a galaxy as a diffuse and dynamically distinct component, which is called a stellar halo. This extended stellar component is thus important to understand the accretion history of a galaxy.

Observations of stellar halos of disk galaxies and studies of their properties have been limited to nearby large galaxies such as the Milky Way and M31. In the Milky Way, the nature of the stellar halo has been explored by wide-field surveys such as SDSS \citep{york2000}, Pan-STARRS \citep{chambers2016}, DES \citep{abbott2018}, {\it Gaia} mission \citep{gaiacollaboration2016}, HSC-SSP \citep{aihara2018a}, and so on. Thanks to these surveys, it has been shown that Milky Way's stellar halo is characterized by dual global halo structures \citep[e.g.,][]{carollo2007,fukushima2018}. In particular, many observations found that the inner halo profile of the Milky Way shows a shallow power-law profile with $\alpha \sim -3$ \citep[e.g.,][]{juric2008}, while its outer halo profile is represented as a steep profile with $\alpha \sim -6$ for the Galactocentric radius of $50 \lesssim r \lesssim 100$ kpc \citep{deason2014a} and again shows a shallow profile with $\alpha = -3.2 \sim -4.0$ at $100 \lesssim r \lesssim 200$ kpc \citep[e.g.,][]{thomas2018,fukushima2019}. Comparison to cosmological simulations suggests that the steep profile beyond $r \sim 50$ kpc indicates that the Milky Way has experienced a relatively quiet accretion history over the past few gigayears \citep{deason2014a}. Besides this, these surveys have also increased the number of newly identified satellite galaxies and substructures \citep[e.g.][]{homma2016,homma2023,suzuki2024}. In M31, the Pan-Andromeda Archaeological Survey \citep[PAndAS;][]{mcconnachie2009} and the SPLASH survey \citep{gilbert2009} have shown that its halo has a power-law surface brightness profile \citep[e.g.][]{ibata2014} and extends to 175 kpc \citep{gilbert2012}. In addition, beyond the Local Group, halos in many large galaxies (e.g. M81, NGC4631, and M101) which are the comparable mass and size as the Milky Way have been observed \citep{okamoto2015,tanaka2017,jang2020}. However, for smaller galaxies such as M33, few observations have been conducted to probe the nature of the stellar halo due to its faintness.

M33 is a dwarf spiral galaxy located at 859 kpc from the Sun \citep{degrijs2017} and it is a possible satellite of M31. M33 has a stellar mass of $3.2 \times 10^{9} M_{\odot}$ \citep{vandermarel2012} and a virial mass of $2 \times 10^{11} M_{\odot}$ \citep{corbelli2014}. Applying the stellar mass-halo mass relation \citep[e.g.,][]{behroozi2019}, M33 should have a stellar halo with a mass of $8 \times 10^{7} M_{\odot}$.

The existence of the M33 halo has been the subject of much debate. PAndAS was unable to detect its stellar halo down to $\mu_V< 35.5$ mag arcsec$^{-2}$ in the region within 3.75 deg from the center of M33 \citep{mcmonigal2016}. In addition, recent narrow-band photometry which was conducted out to a maximum-projected distance of about 40 kpc from the center of M33 has failed to detect any planetary nebulae outside the disk, suggesting the absence of a stellar halo \citep{galera-rosillo2018}. However, recent spectroscopic observations have revealed a dynamically hot population in the central region of M33, which is different from the disk component \citep{gilbert2022}. Moreover, HST multi-color imaging (Panchromatic Hubble Andromeda Treasury:
Triangulum Extended Region survey; PHATTER survey) proposed that M33 has a power-law stellar halo with an index of $\alpha\sim -3$, over the projected radius of $R=5$ kpc from the center of M33 \citep{smercina2023}. However, the studies that find evidence for a stellar halo in M33 have been limited to the inner region ($R<5$ kpc) where the disk population dominates \citep[disk scale length is 1.86 kpc;][]{kam2015}. So, the outer region, which is unique to the stellar halo, including the presence of faint substructures and detection of the edge of the stellar halo, has not been clarified.

In PAndAS, an extended disk was found in M33, suggesting an interaction between M31 and M33 \citep{mcconnachie2009}. It was also confirmed that M33 accounts for $73\%$ of the total stellar mass of M31 satellites \citep{mcconnachie2018}. Therefore, the interaction between M31 and M33 would make a significant contribution to the merger history of M31. \citet{ibata2007} reported that M33 was confirmed as metal-poor and was initially expected that a stellar halo would be detected in PAndAS. However, statistical modeling of M33 failed to detect the stellar halo plausibly because the data were too shallow \citep{mcmonigal2016}.

To investigate the global properties of the M33 stellar halo, we have been conducting a photometric survey of M33 using Subaru/Hyper Suprime-Cam (HSC). HSC is an outstanding instrument for detailed and high-quality detection and characterization of the M33 stellar halo using the faint red giant branch (RGB), red clump (RC), and horizontal branch (HB) stars as tracers. HSC's three broadband filters, ${\it g/r_2/i_2}$-bands, and one narrowband filter, {\it NB515}, are a good combination to remove contaminants. By combining the three broadband filters, it is possible to remove unresolved background galaxies using color selections. {\it NB515}, centered on the MgH/Mgb lines which are sensitive to stellar surface gravity \citep{majewski2000}, can be used to remove foreground Galactic dwarf stars \citep{komiyama2018,ogami2024a}. Therefore, the removal of these contaminants and the wide and deep HSC observations are expected to provide the nature of the faint M33 stellar halo.

In this paper, we report the initial results of the ongoing survey, and analyses of the M33 halo using the Subaru/HSC. We focus on the western region of M33, which is perpendicular to the major-axis of the disk and is not significantly affected by contamination from the extended disk of M33, to investigate the properties of the stellar halo. From this analysis, we demonstrate the superiority of Subaru/HSC for detecting the faint stellar halo in M33. The paper is organized as follows. In Section \ref{section:obs}, we describe our observational data and data reduction using the HSC pipeline. In Section \ref{section:selection}, we introduce the method to select the halo stars of M33. The results of the radial profiles of M33 are given in Section \ref{section:profile}, and the discussion for these results is presented in Section \ref{section:Discussion}. Finally, Section \ref{section:Conclusion} concludes this paper.

\begin{deluxetable*}{ccccc}
\tablecaption{The details of our observations.\label{tab:obs}}
\tablewidth{0pt}
\tablehead{
\colhead{Band} & \colhead{Date} & \colhead{Exposure Time} & \colhead{Seeing FWHM} & \colhead{50 \% detection completeness magnitude}\\
& \colhead{[mm/yyyy]} & \colhead{[seconds]} & \colhead{[arcsec]} & \colhead{[mag]}}
\startdata
${\it g}$-band & 10/2022 & $60\times5 + 300\times15$ & $0.54\pm0.03$ & 26.63\\
${\it r_2}$-band & 10/2022 \& 01/2023 & $60\times5 + 300\times45 + 340\times4$ & $0.90\pm0.13$ & 26.10\\
${\it i_2}$-band & 10/2022 & $60\times5 + 204\times30$ & $1.19\pm0.16$ & 25.05\\
${\it NB515}$-band & 09/2023 & $240\times 35$ & $0.83\pm0.14$ & 24.21\\
\enddata
\end{deluxetable*}

\begin{figure}[ht!]
\includegraphics[width=\columnwidth]
{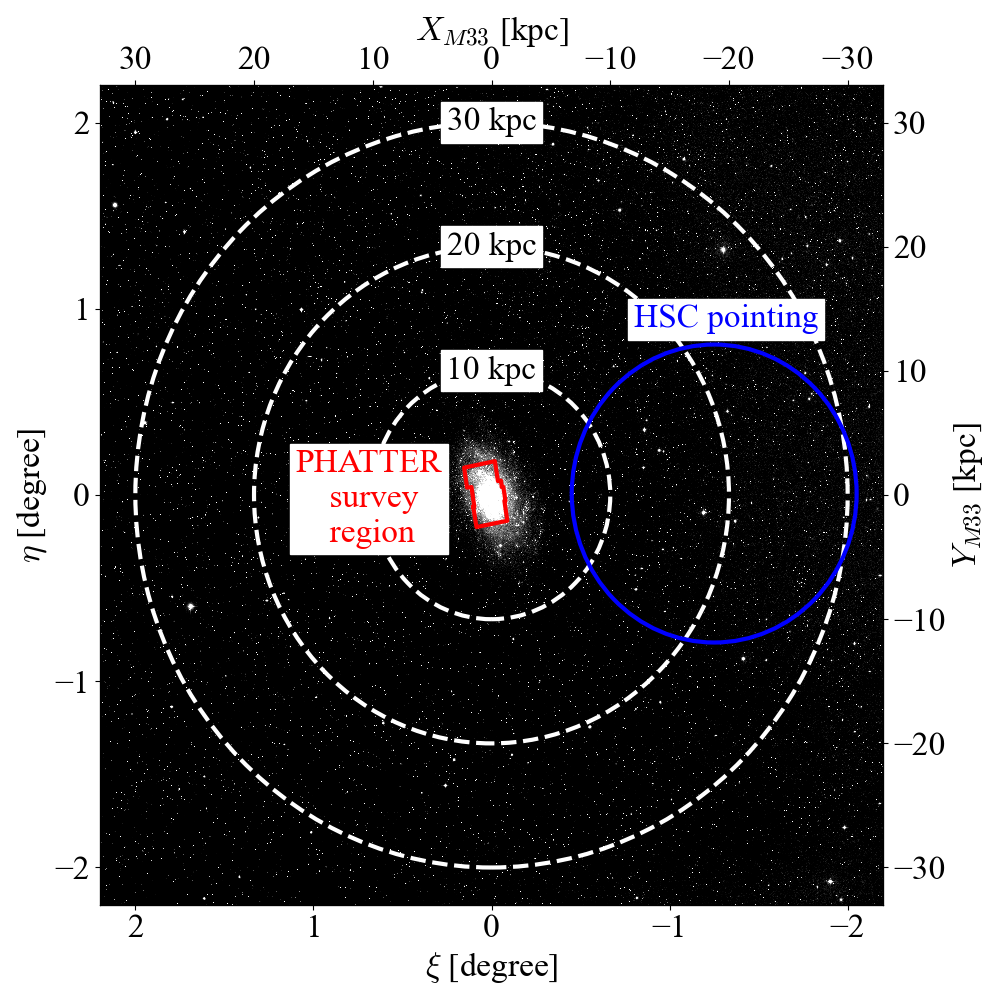}
\caption{The blue circle indicates the HSC observation field showing a tangent plane centered on M33. The red polygon shows the survey fields observed by the Hubble Space Telescope (PHATTER survey). The background image is taken from Pan-STARRS1 \citep{schlafly2012,tonry2012,magnier2013,flewelling2020}.
\label{figure:ObsMap}}
\end{figure}

\section{Observation and Data Reduction} \label{section:obs}
We observed the western region of M33 in the \textit{g}-, \textit{$r_2$}- and \textit{$i_2$}-bands and \textit{NB515} using Subaru/HSC during the nights of 2022 and 2023 (PI: I. Ogami; Proposal ID: S22B-107 \& S23B-072). The observed region is the single pointing, which corresponds to the field-of-view of HSC ($\sim$ 1.76 degree$^2$), shown as the blue circle in Figure \ref{figure:ObsMap}. In Figure \ref{figure:ObsMap}, we show the observational fields of the PHATTER survey \citep{williams2021,smercina2023} as a reference. Our field reaches a projected radius up to $\sim 30$ kpc from the center of M33. In this outer region, the disk contamination, which includes the perturbed disk structure \citep{mcconnachie2009}, is relatively low, because this region is perpendicular to the major-axis of the M33 disk and the orientation of the extended disk, so we can focus on the presence of the stellar halo. Our observations were carried out under relatively good seeing conditions (0\farcs50 to 1\farcs1).

The observed raw images are processed and calibrated using the HSC pipeline \citep[hscPipe version 8.4;][]{bosch2018}. The hscPipe is based on software for the Large Synoptic Survey Telescope \citep[LSST;][]{ivezic2008} project. The hscPipe conducts data reduction, including bias and dark subtraction, flat fielding, sky subtraction, and cosmic ray removal. After the reduction for individual CCDs, this pipeline calibrates the coordinates and flux scales using Pan-STARRS1 \citep[PS1;][]{schlafly2012,tonry2012,magnier2013,flewelling2020}, stacks each frame, and performs the source detection and photometry to create the scientific catalog. After the reduction and calibration, to evaluate the detection completeness, we perform artificial star tests using the hscPipe and \texttt{injectStar.py} \citep{ogami2024a}. We conduct the same method as in \citet{ogami2024a}. Briefly, we embed the artifitial stars with given magnitudes into the reduced images using \texttt{injectStar.py}, then we conduct the detection and photometry using the hscPipe. The outcome of this test (see, Table \ref{tab:obs}) indicates that our broad-band photometry achieves over 1 magnitude deeper than that of the previous survey \citep[PAndAS;][]{ibata2014,mcconnachie2018}. 

Based on the Galactic dust extinction \citep{schlegel1998,schlafly2012}, we apply the extinction correction for each source. The extinction-corrected magnitudes for each band are
\begin{align}
\textit{g}_0 &= \textit{g} - 3.676\times E(B-V) \notag \\
\textit{r}_{2,0} &= \textit{r}_2 - 2.584\times E(B-V)\\
\textit{i}_{2,0} &= \textit{i}_2 - 1.903\times E(B-V) \notag\\
\textit{NB515}_0 &= \textit{NB515} - 2.862\times E(B-V) \notag
\end{align}
where the subscript "$_0$" means the extinction-corrected magnitude. The coefficient of the extinction-corrected magnitude of each band is obtained by the same method of \citet{ogami2024a} which multiplies the interstellar absorption curve of \citet{fitzpatrick1999} with $R=+3.1$ by the SED of a G-type star (${\rm T_{eff}}=7000$ K, $\log{\rm Z}=-1$, $\log{g}=4.5$) and integrates it using the response curve of each band. 

\begin{figure}[ht!]
\includegraphics[width=\columnwidth]
{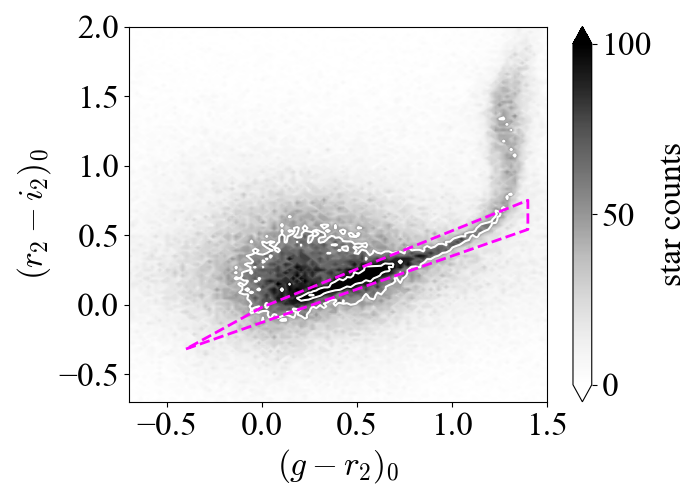}
\caption{The $({\it g}-{\it r_2})_0$ v.s. $({\it r_2}-{\it i_2})_0$ space of all point sources, which are determined by the \texttt{extendedness} parameter of $\it{g}$-band. The white contours represent where the star counts are 100 and 200. In the following analysis, we regard the objects within the purple polygon as the `pure' stellar point sources.
}\label{figure:CCD}
\end{figure}

\section{Selection of the M33 Halo Stars}\label{section:selection}
\subsection{Point Sources}\label{section:pointsources}
In the hscPipe, the parameter, \texttt{extendedness} which determines whether the object is a point source (\texttt{extendedness == 0}) or an extended source (\texttt{extendedness == 1}), is constructed for each band. In the following analysis, we mainly use those objects that are determined to be the point sources in {\it g}-band, because the seeing of ${\it g}$-band images is the best of all-bands in our study (seeing $\sim 0\farcs5$). However, \citet{aihara2018} reported that it is difficult to perform star-galaxy classification using the \texttt{extendedness} parameter fainter than $24.5$ mag of {\it i}-band. To overcome this problem, we use the $({\it g}-{\it r_2})_0$ vs. $({\it r_2}-{\it i_2})_0$ diagram to extract `pure' stellar point sources, in addition to the \texttt{extendedness} parameter. Figure \ref{figure:CCD} shows the color-color diagram of the point sources which are determined by the \texttt{extendedness} parameter of ${\it g}$-band. In this space, it is known that the stellar sequence occupies the region from $(-0.2, -0.2)$ to $(1.3, 0.5)$ \citep[e.g.,][]{lenz1998,krisciunas1998}. This sequence bends around $(({\it g}-{\it r_2})_0,({\it r_2}-{\it i_2})_0) \sim (1.3,0.5)$, because there are the many foreground M- and K-type stars. In this space, there are also unresolved background objects around $(({\it g}-{\it r_2})_0,({\it r_2}-{\it i_2})_0) \sim (0.2, 0.2)$, which deviates from the stellar sequence. These objects are composed of quasars and star-forming galaxies, as confirmed by spectroscopic observations \citep{ahumada2020,lyke2020}. Therefore, in this study, we regard the objects within the purple polygon which is drawn by dashed lines in Figure \ref{figure:CCD} as the `pure' stellar point sources. This purple polygon is almost the same as \citet{suzuki2024}, and covers the PARSEC isochrones \citep{bressan2012,marigo2017} with $-0.5<[{\rm Fe/H}]<-2.5$ and $8~{\rm Gyr} < {\rm Age} < 13~{\rm Gyr}$. We use these `pure' stellar point sources for the following analysis. Finally, our stellar sample does not include cooler stars (e.g., K and M-type stars), so our sample is a biased selection. However, this bias is probably fine, because we think that the halo of M33 should be as metal-poor as that of the large galaxies (e.g., the Milky Way).

\begin{figure}[ht!]
\includegraphics[width=\columnwidth]
{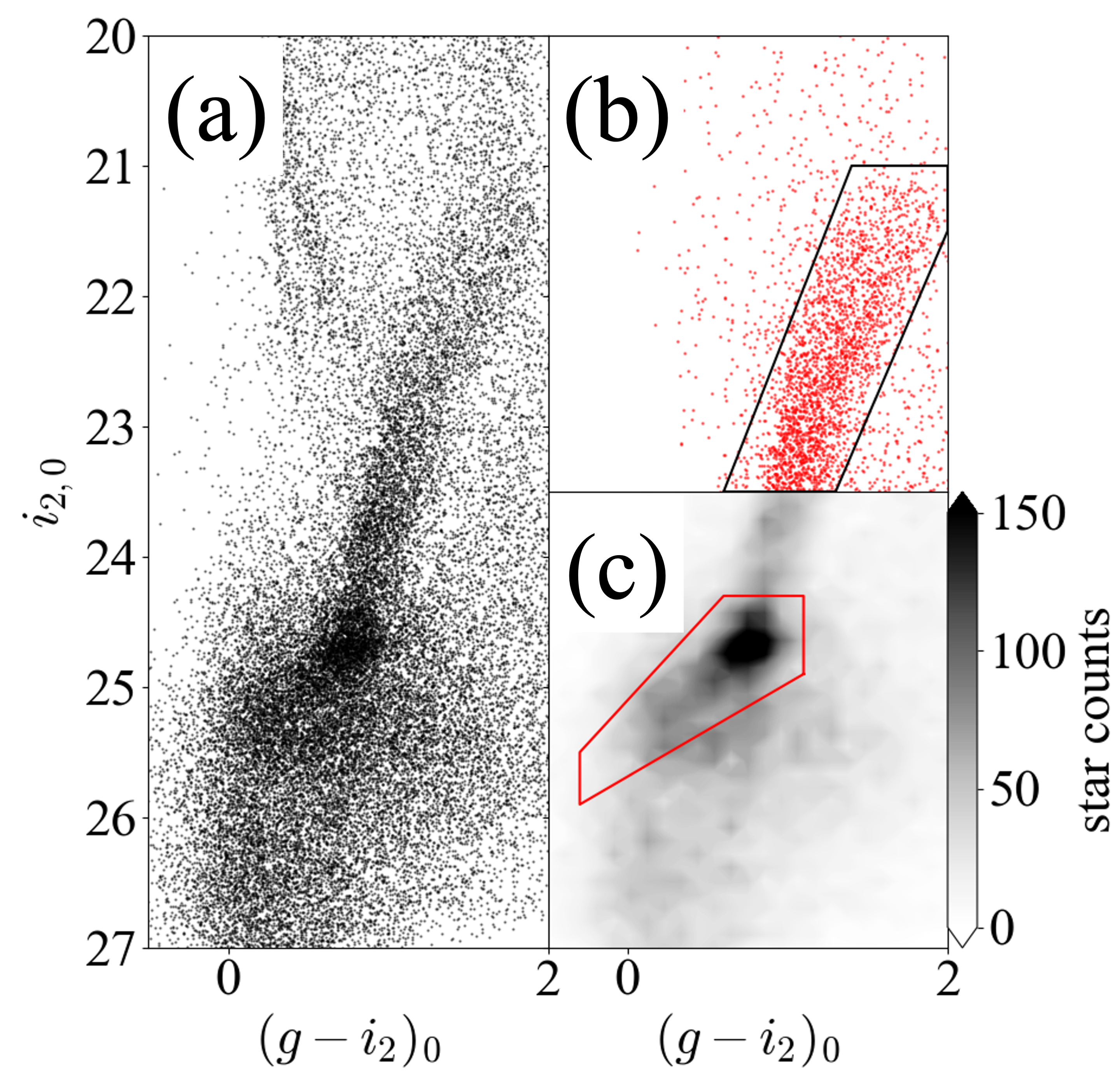}
\caption{(a) The color-magnitude diagram of the `pure' stellar point sources which is determined by the $({\it g}-{\it r_2})_0$ v.s. $({\it r_2}-{\it i_2})_0$ space. (b) The color-magnitude diagram of the `pure' stellar point sources with $p_{RGB}>0.9$. We consider stars in the black polygon as {\it NB515}-selected Red Giant Branch (NRGB) stars for our analysis. (c) The same as in panel (a) represented by a Hess Diagram. We consider stars in the red polygon as Red Clump stars for our analysis.
\label{figure:CMD}}
\end{figure}

\subsection{RGB and RC Stars in M33}\label{section:RGB_RC}
Figure \ref{figure:CMD}(a) shows the color-magnitude diagram (CMD) of all `pure' stellar point sources. In this diagram, we can see the clear RGB sequence from $(({\it g}-{\it i_2})_0,{\it i_{2,0}}) \sim (2,21)$ to $(1,24)$. However, due to the low galactic latitude of M33, foreground dwarf stars can be also seen from $\sim (2,20)$ to $(2,24)$ (Galactic disk stars) and from $\sim (0,21)$ to $(1,23)$ \citep[Galactic halo substructures;][]{martin2014} in this diagram. Due to the overlapping of the foreground contaminants with the M33 RGB stars, it is difficult to separate the M33 halo stars from the foreground stars using only the color-magnitude diagram. To solve this problem, \citet{ogami2024a} introduced the RGB probability, $p_{\rm RGB}$, which is the probability that the star is on the RGB calculated by the {\it NB515} information. Briefly, {\it NB515} is a narrow-band filter covering the MgH band and the Mgb triplet which are the surface gravity-sensitive absorption lines, so the dwarf stars and RGB stars are distributed in different locii on the $({\it NB515}-{\it g})_0$ v.s. $({\it g}-{\it i_2})_0$ diagram \citep[see Figure 6 in][]{komiyama2018}. Using this difference on the color-color diagram, \citet{ogami2024a} constructed a probability distribution function of dwarf stars, and calculated the dwarf probability and RGB probability based on this information. In addition to {\it NB515}-information, they also took into account the galactic latitude information in the final RGB probability. However, since the galactic latitude does not change significantly in our observation region, we derive the RGB probability only using {\it NB515}-information. We calculate $p_{\rm RGB}$ for the stars with ${\it i_{2,0}} < 23.5$, and the CMD constructed by the stars with $p_{\rm RGB}>0.9$ is shown in Figure \ref{figure:CMD}(b). \citet{ogami2024a} reported that even if $p_{\rm RGB}$ was possible to remove contamination with $\sim 90$ \% accuracy, there were a few remaining foreground stars with $(g-i_2)_0 \sim 2.5$. Therefore, we extract the robust RGB stars by selecting stars with $p_{\rm RGB}>0.9$ and also by selecting stars within the black polygon in Figure \ref{figure:CMD}(b). \citet{ogami2024a} applied the color-cut at $({\it g} - {\it i_2})_0 \sim 2.5$, but this study applies the color-cut at $({\it g}-{\it i_2})_0 = 2.0$ in the CMD. This is because we have already conducted the color-cut for the selection of `pure' point sources at $({\it g}-{\it r_2})_0 \sim 1.5$ (see Section \ref{section:pointsources}), so our sample has few stars with $2<({\it g} - {\it i_2})_0<2.5$. In the following analysis, the objects in the black polygon on Figure \ref{figure:CMD}(b) are called {\it NB515}-selected RGB (NRGB) stars.

Figure \ref{figure:CMD}(c) is the Hess diagram of all `pure' stellar point sources, and we can confirm the overdensity around $i_{2,0} \sim 24.5$. This overdensity corresponds to the Red Clump (RC), assuming a typical distance of M33 \citep[859 kpc;][]{degrijs2017}. In general, RC stars are known to be more numerous than RGB stars, so RC stars are advantageous for studying low-surface brightness structures such as the stellar halo. Therefore, in this study, we consider the `pure' point sources within the red polygon, which covers the PARSEC isochrones with $-0.5<[{\rm Fe/H}]<-2.5$ and $8~{\rm Gyr} < {\rm Age} < 13~{\rm Gyr}$ in Figure \ref{figure:CMD}(c) as RC stars. Using these RC stars, we independently probe the stellar halo of M33 and search with NRGB stars. Incidentally, it should be noted that we do not use {\it NB515}-information to select our RC samples, because RC stars are difficult to separate from foreground dwarf stars in {\it NB515} color space due to overlapping with the foreground stars.

\section{Radial Profile}\label{section:profile}

\begin{figure*}[ht!]
\includegraphics[width=2\columnwidth]
{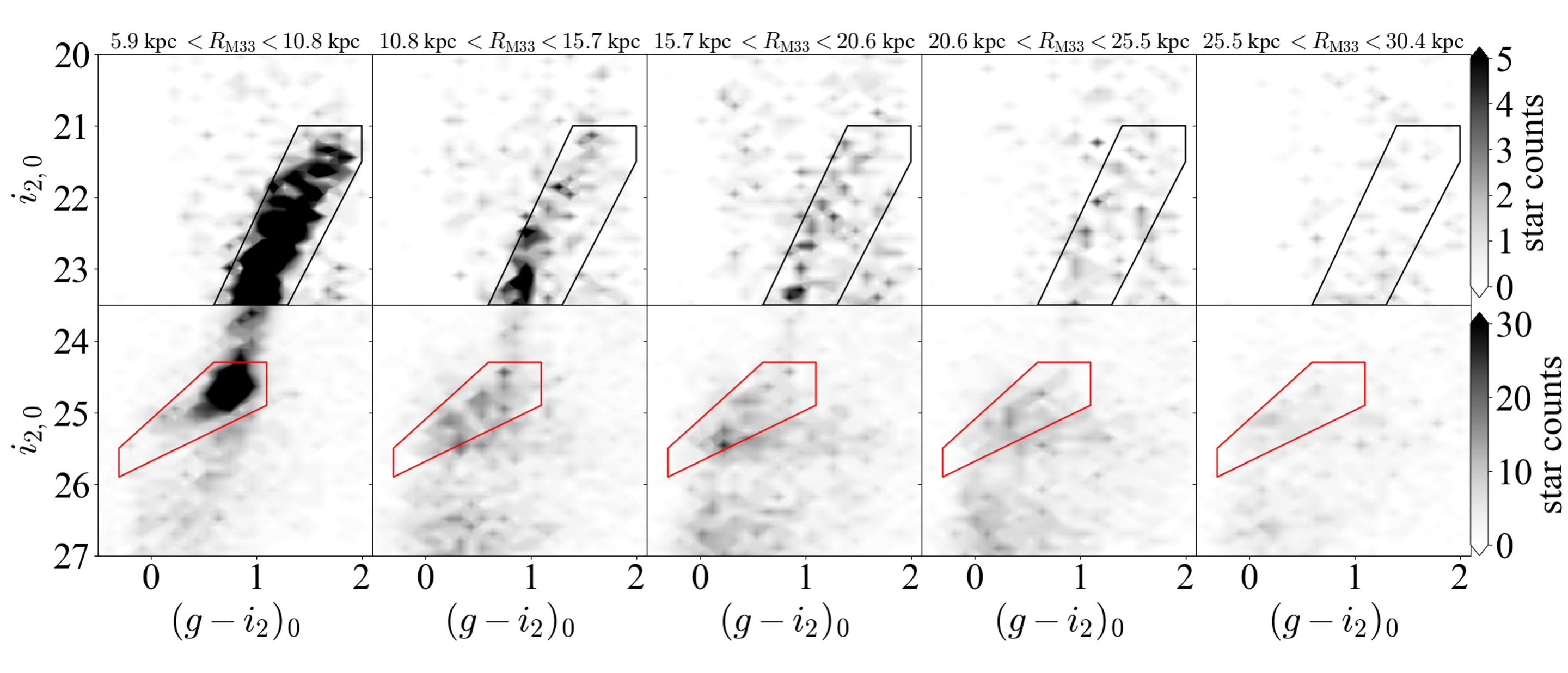}
\caption{Top-row: The Hess diagrams of `pure' stellar point sources with $p_{\rm RGB}>0.9$. The black polygons show the boundary of the RGB defined in Section \ref{section:RGB_RC}. Bottom-row: The Hess diagrams of `pure' point sources. The red polygons show the boundary of the RC defined in Section \ref{section:RGB_RC}. It is noted that the color scale representing the number density is different in the top and bottom-rows. 
\label{figure:CMD_region}}
\end{figure*}

First, we construct the Hess diagrams for each region using the `pure’ point sources defined in Section \ref{section:selection} as shown in Figure \ref{figure:CMD_region}. The divided area is a region that roughly divides the HSC's field of view into five equal areas. The CMDs in the top-row are constructed by `pure’ point sources with $p_{\rm RGB} > 0.9$, and those in the bottom-row are constructed by `pure' point sources. The boxes for the RGB and RC stars defined in Section \ref{section:RGB_RC} are also shown in this figure with black and red boxes, confirming that our selected color-magnitude cuts adequately cover the RGB and RC in M33. As seen in these panels, we can confirm the presence of RGB sequences and RC stars beyond 10 kpc from the center of M33. In this section, we construct two types of radial profiles to unravel the information obtained from these Hess diagrams in detail.

\subsection{Radial Density Profile}\label{section:DensityProfile}

\begin{figure*}[ht!]
\includegraphics[width=2\columnwidth]
{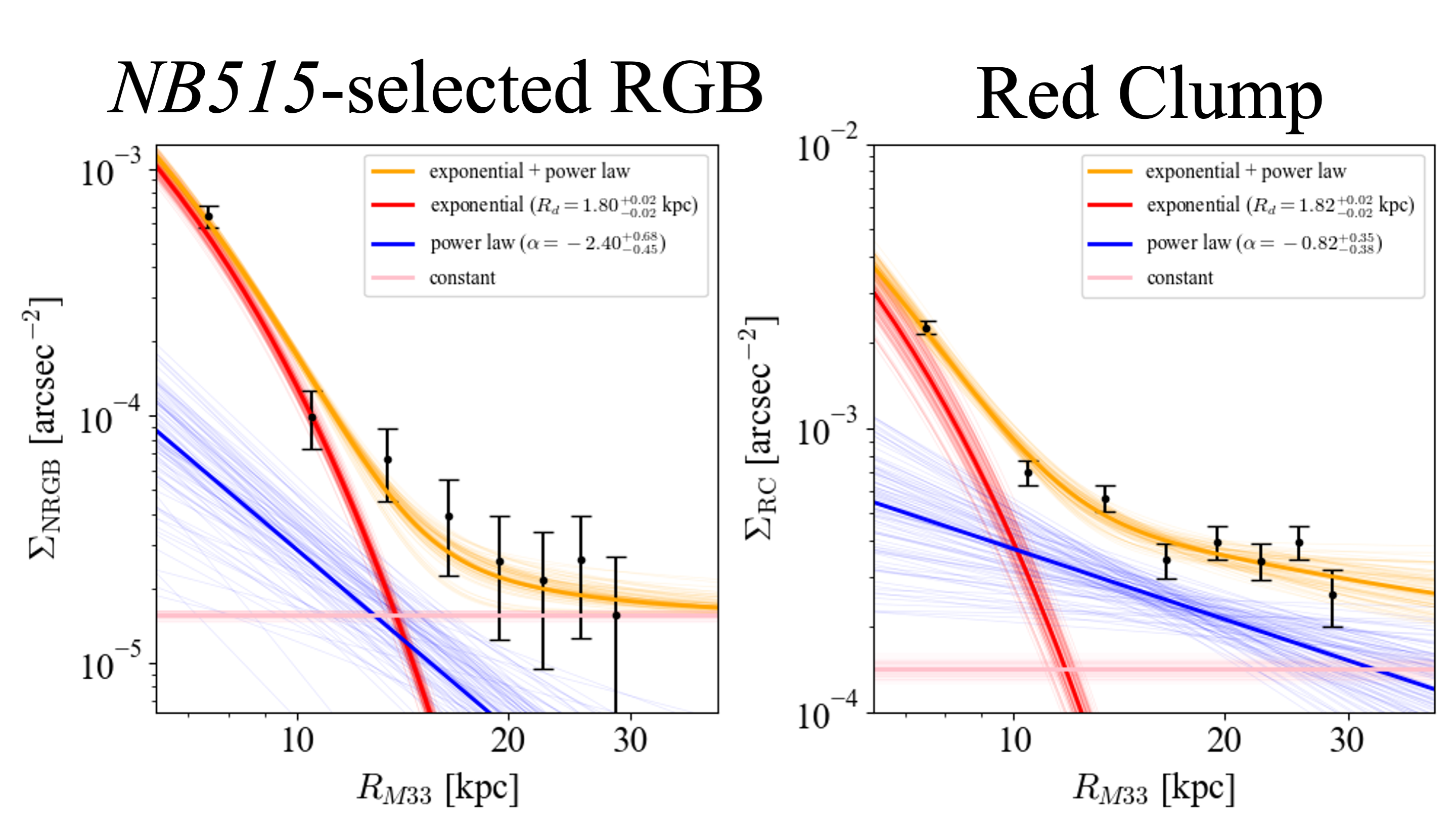}
\caption{Left: The radial density profile of {\it NB515}-selected RGB (NRGB) stars. The black dots and their error bars indicate the number density of stars in each region and their Poisson errors. The orange line shows the MCMC-fitted profile result (halo, disk plus remaining contamination components). The red, blue, and pink lines are the individual profiles for the disk, halo, and contamination components, respectively, when we assume a prior distribution of $R_d$ as a Gaussian distribution with 1.86 kpc. For each line, the thick line shows the profile using median values of the parameters estimated by MCMC and thin lines indicate the results of 1,000 random samples from the posterior distribution. Right: The same as in the left panel, but the profile is constructed using RC. The constant contamination population is not plotted in this figure due to its small value.
\label{figure:profile}}
\end{figure*}

\begin{figure*}[ht!]
\includegraphics[width=2\columnwidth]
{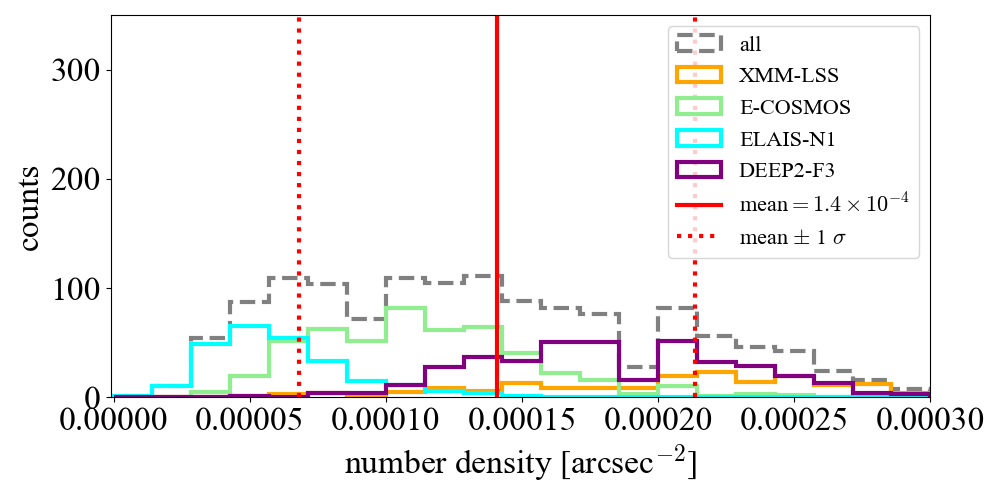}

\caption{The histogram of the surface number density of remaining background galaxies from the HSC-SSP DEEP data. The gray dashed line shows a histogram using all the data in the HSC-SSP DEEP region, and the solid and dashed red lines are the mean and 1-$\sigma$-value of this histogram. The orange, light-green, cyan, and purple lines show histograms for four regions comprising the HSC-SSP DEEP data (XMM-LSS, E-COSMOS, ELAIS-N1, and DEEP2-F3).
\label{figure:SSP}}
\end{figure*}

We construct radial density profiles using the NRGB stars and RC stars defined in Section \ref{section:selection} by counting the number of stars in a given region, as follows. First, we limit the analyzed region to 0.5 degree width along the minor axis of the M33 disk ($-0.25 < \eta < 0.25$ in Figure \ref{figure:ObsMap}). Second, this region is divided into small regions by 0.2 degree along the minor-axis of the M33 disk (i.e., $\xi$ direction). Third, we count the stars within each divided small region. Then, we take into account the detection completeness. Especially for faint stars such as RC stars, there is some loss of observational data due to detection completeness. To compensate for this, when we count stars of a given magnitude, we correct for detection completeness in the radial profile multiplying by the inverse of the detection completeness at that magnitude. Finally, the radial density profile is constructed by dividing the sum of the stellar counts by the area of the region. The constructed profiles of NRGB stars and RC stars are shown with black points in Figure \ref{figure:profile}.

It is known that the limiting magnitude of HSC becomes slightly shallower at the edge of the field of view \citep[$\sim 0.1$ degree from the edge][]{aihara2018}. In our data, the $5\sigma$ limiting matnigude in each band is also $\sim 0.05$ mag shallower on the periphery. Therefore, to perform the robust analysis, we exclude the innermost and outermost data, which are located at the edges of the field of view, and Figure \ref{figure:profile} shows the radial profile after excluding these data points.

To examine the contribution of the disk and halo populations, we fit models to these profiles. We consider a three-component model, with an exponential disk, a power-law stellar halo, and a uniform component assuming remaining contamination (foreground stars for the profile of NRGB stars and background galaxies for the profile of RC stars):
\begin{align}
\Sigma(R) = \Sigma_{\rm disk,0}\exp{\left(-\frac{R}{R_{\rm d}}\right)} + \Sigma_{\rm halo,0}\left(\frac{R}{\rm kpc}\right)^{\alpha} + \Sigma_{\rm c},
\label{eq:profile}\end{align}
where $R_{\rm d}$ is the disk scale length, $\alpha$ is the projected power-law index for the stellar halo, and $\Sigma_{\rm disk,0}$, $\Sigma_{\rm halo,0}$ and $\Sigma_{\rm c}$ are the surface density scale factors for the disk, halo, and contamination, respectively. Based on this equation, we construct a likelihood function to fit the binned radial density profile (see, Figure \ref{figure:profile});
\begin{align}
\begin{split}
\mathcal{L} = \prod_{n} &\frac{1}{\sqrt{2\pi \sigma_i^2}}\\
&\exp{\left[ \frac{-(y_i - \Sigma(R_i|\Sigma_{\rm disk,0},R_{\rm d},\Sigma_{\rm halo,0},\alpha,\Sigma_{\rm c}))^2}{2\sigma_i^2}\right]},    
\end{split}
\end{align}
where $\Sigma_{\rm disk,0}$, $R_{\rm d}$, $\Sigma_{\rm halo,0}$, $\alpha$, and $\Sigma_{\rm c}$ are the same parameter in Equation (\ref{eq:profile}). $R_i$ and $y_i$ correspond to the radial distance and stellar density of the $i$-th data point of the radial profile shown in Figure \ref{figure:profile}, respectively. $\sigma_i$ is the uncertainty of the stellar density. In the case of the profile of the NRGB stars, we use wide and flat prior distributions (i.e., uninformed prior distributions) for $\Sigma_{\rm disk, 0}$, $\Sigma_{\rm halo, 0}$, $\alpha$, and $\Sigma_{\rm c}$. It is difficult to estimate the disk properties because our observation region covers the outside of the typical M33 disk. Hence we adopt a Gaussian distribution with a mean of 1.86 kpc as the disk scale length $R_d$ and its standard deviation of 0.02 kpc, which is the statistical uncertainty of $R_d$, from \citet{kam2015}, which was estimated by the Spitzer/IRAC $3.6~{\rm \mu m}$ data. In addition to an assumed Gaussian distribution with a mean of 1.86 kpc, we adopt a Gaussian distribution with a mean of 4.34 kpc, because the disk scale length of M33 was estimated to be 4.34 kpc in a previous study \citep{smercina2023}, which detected the stellar halo of M33.

In the case of the RC stars, we also use the same uniform priors as the RGB stars for $\Sigma_{\rm disk,0}$, $\Sigma_{\rm halo,0}$ and $\alpha$, and the same Gaussian distributions as the RGB stars for $R_{\rm d}$. However, for $\Sigma_{\rm c}$, we adopt a Gaussian distribution of the mean and standard deviation values derived from observational data instead of an uninformed prior, because it is possible to use observational data to estimate the contribution from background galaxies. The amount of remaining background galaxies is estimated using data from the HSC-SSP DEEP region \citep{aihara2018a}. We conduct the same reduction and calibration as our observational data, and we perform the same color and magnitude cut for the comparison data (i.e., for HSC-SSP DEEP data, we select the objects within the purple polygon in Figure \ref{figure:CCD} and the red polygon in Figure \ref{figure:CMD} (c)). After that we count the number of sources for each $0.5 \times 0.2$ degree$^2$ region for these data correcting its detection completeness. Figure \ref{figure:SSP} shows the histogram of the number density of remaining background galaxies from the HSC-SSP DEEP data. The HSC-SSP DEEP consists of observations for four different regions. Therefore, Figure \ref{figure:SSP} shows the distribution of all HSC-SSP DEEP data as gray dashed lines, and the distribution of each region as orange, light-green, cyan, and purple solid lines. From this figure, we find that the surface density of remaining objects in the HSC-SSP DEEP data is a mean value of $1.4\times 10^{-4}$~arcmin$^{-2}$ and a standard deviation of $7.3\times 10^{-5}$~arcmin$^{-2}$, shown as red solid and dashed lines in Figure \ref{figure:SSP}. We set the Gaussian prior for the parameter $\Sigma_{\rm c}$ using these mean and standard deviation values. From Figure \ref{figure:SSP}, we see that the number of remaining background objects varies from region to region, and the standard deviation is large relative to the mean value. The identification of the diffuse structure, such as the stellar halo, is particularly sensitive to the background contamination level due to the low expected number density of the structure. Therefore, we also fit profiles with fixed $\Sigma_{\rm c}$ (not as a parameter) with the maximum or minimum value obtained from HSC-SSP data. 

\begin{figure*}[ht!]
\includegraphics[width=2\columnwidth]
{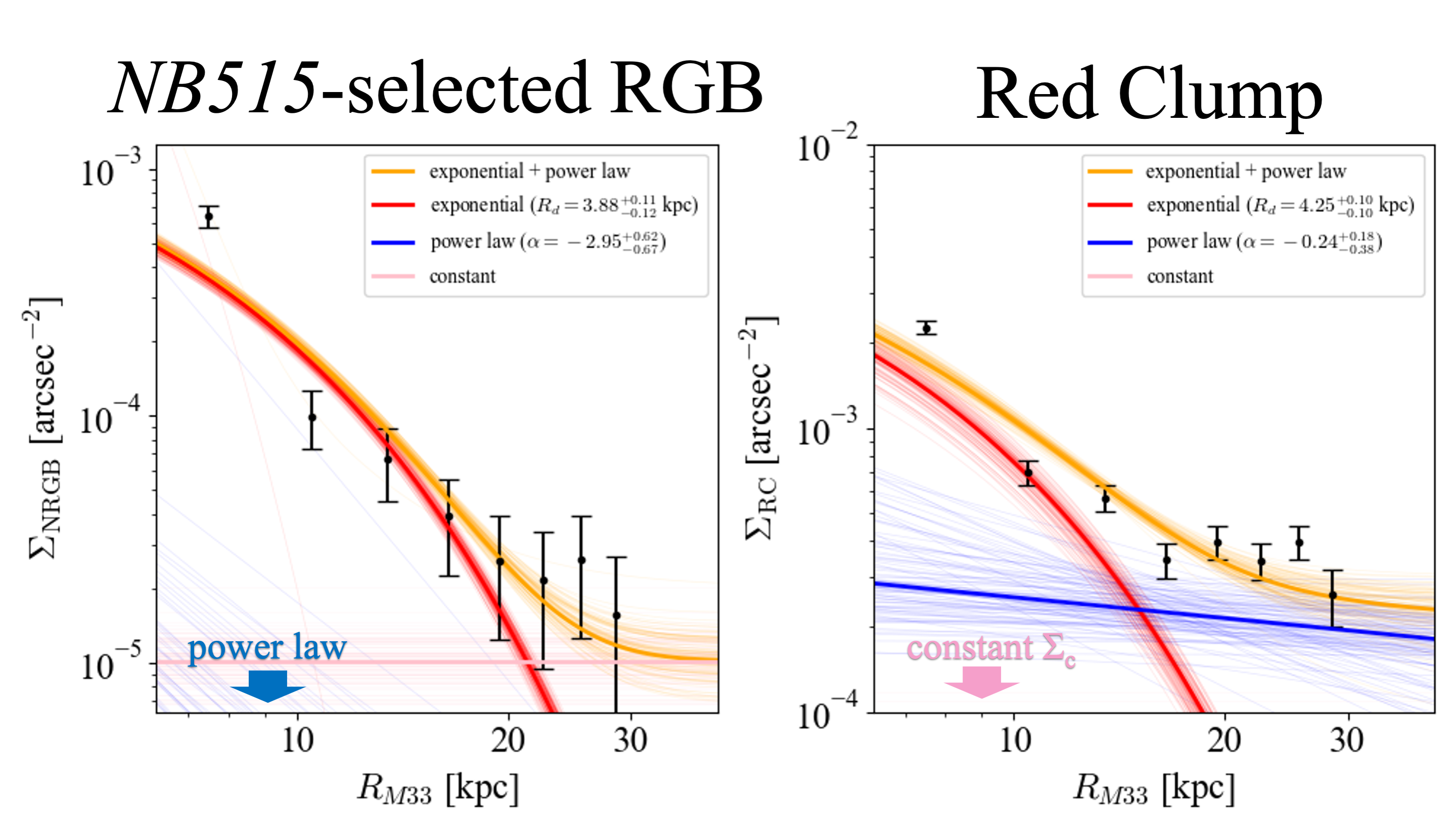}
\caption{The same as in Figure \ref{figure:profile} but an assumed prior for $R_d$ is a Gaussian distribution with a mean of 4.34 kpc.
\label{figure:profile_434}}
\end{figure*}

\begin{deluxetable*}{cccccc}[ht!]
\tablecaption{The estimated results of the radial profiles. These values are the median of the posterior distribution and the 68\% credible intervals.\label{table:profile}}
\tablecolumns{6}
\tablewidth{0pt}
\tablehead{
\colhead{Case}&
\colhead{$\Sigma_{\rm disk, 0}$} &
\colhead{$R_{\rm d}$} & 
\colhead{$\Sigma_{\rm halo, 0}$} &
\colhead{$\alpha$} &
\colhead{$\Sigma_{\rm c}$}\\
\colhead{}& 
\colhead{$[{\rm arcsec}^{-2}]$} &
\colhead{[${\rm kpc}]$} & 
\colhead{[${\rm arcsec}^{-2}]$} &
\colhead{}&
\colhead{[${\rm arcsec}^{-2}]$}
}
\startdata
NRGB ($R_d=1.86$ kpc, Uniform prior) & $0.034_{-0.003}^{+0.003}$ &  $1.80_{-0.02}^{+0.02}$  & $0.007_{-0.0006}^{+0.014}$ & $-2.40_{-0.45}^{+0.68}$ & $0.0000157_{-0.0000004}^{+0.0000004}$\\
NRGB ($R_d=4.34$ kpc, Uniform prior) & $0.0024_{-0.0002}^{+0.0002}$ &  $3.88_{-0.11}^{+0.11}$  & $0.0014_{-0.0009}^{+0.0008}$ & $-2.95_{-0.67}^{+0.62}$ & $0.000010_{-0.000002}^{+0.000002}$\\
RC ($R_d=1.86$ kpc, Gaussian prior) & $0.10_{-0.01}^{+0.01}$ &  $1.82_{-0.02}^{+0.02}$ & $0.0025_{-0.001}^{+0.005}$ & $-0.82_{-0.37}^{+0.35}$ & $0.00014_{-0.00001}^{+0.00001}$\\
RC ($R_d=4.34$ kpc, Gaussian prior) & $0.008_{-0.001}^{+0.001}$ & $4.25_{-0.10}^{+0.10}$ & $0.002_{-0.001}^{+0.003}$ & $-0.92_{-0.37}^{+0.61}$ & $0.00014_{-0.00001}^{+0.0001}$\\
RC ($R_d=1.86$ kpc, maximum $\Sigma_{\rm c}$) & $0.10_{-0.01}^{+0.01}$ & $1.82_{-0.02}^{+0.02}$ & $0.04_{-0.03}^{+0.03}$ & $-2.2_{-0.3}^{+0.5}$ & ($2.9\times10^{-4}$; fixed)\\
RC ($R_d=4.34$ kpc, maximum $\Sigma_{\rm c}$) & $0.008_{-0.001}^{+0.001}$ & $4.22_{-0.01}^{+0.01}$ & $0.004_{-0.003}^{+0.003}$ & $-2.9_{-0.7}^{+0.7}$ & ($2.9\times10^{-4}$; fixed)\\
RC ($R_d=1.86$ kpc, minimum $\Sigma_{\rm c}$)  & $0.10_{-0.01}^{+0.01}$ & $1.82_{-0.02}^{+0.02}$ & $0.001_{-0.001}^{+0.001}$  & $-0.46_{-0.22}^{+0.21}$ & ($1.5\times10^{-5}$; fixed)\\
RC ($R_d=4.34$ kpc, minimum $\Sigma_{\rm c}$)  & $0.077_{-0.0009}^{+0.0007}$ & $4.34_{-0.02}^{+0.02}$ & $0.0004_{-0.0002}^{+0.0007}$  & $-0.19_{-0.31}^{+0.14}$ & ($1.5\times10^{-5}$; fixed)\\
\enddata
\end{deluxetable*}
We perform the MCMC fitting using the Python module \texttt{emcee} \citep{foreman-mackey2013} to examine the contribution of the disk and halo populations. We initialize the sampler with 100 walkers, with 10,000 steps after a burn-in period of 100,000 steps to ensure a good sampling of the posterior distributions. Figure \ref{figure:profile} also shows the results of the profile fitting shown as the thick orange line and the results of 1,000 random samples from the posterior distribution shown as thin orange lines. This figure shows the fitting results when the prior distribution of $R_d$ is assumed to be a Gaussian distribution with a mean of 1.86 kpc. Figure \ref{figure:profile_434} also shows the fitting results when the prior distribution of $R_d$ is assumed to be a Gaussian distribution with a mean of 4.34 kpc. In both figures, the fitting results for each population are indicated by red lines for disk, blue lines for halo, and pink lines for contamination. Table \ref{table:profile} also shows the estimation results for all fitting cases. We show the posterior and marginalized distributions for all fitting cases in Appendix \ref{section:MCMC}.

In the case of RC stars, we confirm that the halo population is sensitive to the amount of contamination. Hence, the value of the alpha index is inconsistent, especially for the cases with the highest and lowest amounts of contamination. Besides this, the fitted RC profiles tend to be shallower than the result of NRGB stars. Taking the statistical error into account, the result of RC stars with the highest contamination is consistent with the result of NRGB stars, so it is possible that there is still some remaining contamination in our RC sample and we may overestimate the M33 halo stars. It is noted that the RC and NRGB stars may be probing slightly different samples. The NRGB stars are biased towards lower metallicities due to the color-magnitude cut, while the RC stars tend to be younger or more metal-rich than the NRGB stars, so this difference in population may affect differences in the results of the radial density profile. 

Our result shows the alpha indices of the halo component are shallower than $-3$, depending on the intensity of contamination sources. Except for the results in the NRGB cases, and the result in the RC case assuming maximum contamination sourses, our estimated values are inconsistent with \citet{smercina2023}, if the stellar halo extends to the outer region with a same power-law with $\alpha \sim -3$ which is estimated by Smercina et al. (2023). However, in this study, we analyze the outer region, while the region where \citet{smercina2023} estimated is the inner region ($<$ 5 kpc; see the red polygon in Figure \ref{figure:ObsMap}). In the Milky Way, the halo profile has a break, and the gradient of the profile changes at the break. Therefore, this result may indicate that the gradient of the stellar halo of M33 changes in the outer region as well; M33 may have double halos (inner and outer halo). Our results that show the shallower population are discussed in detail in Section \ref{section:OuterHalo}.

\subsection{Color Profile}\label{section:ColorProfile}

\begin{figure*}[ht!]
\includegraphics[width=2\columnwidth]
{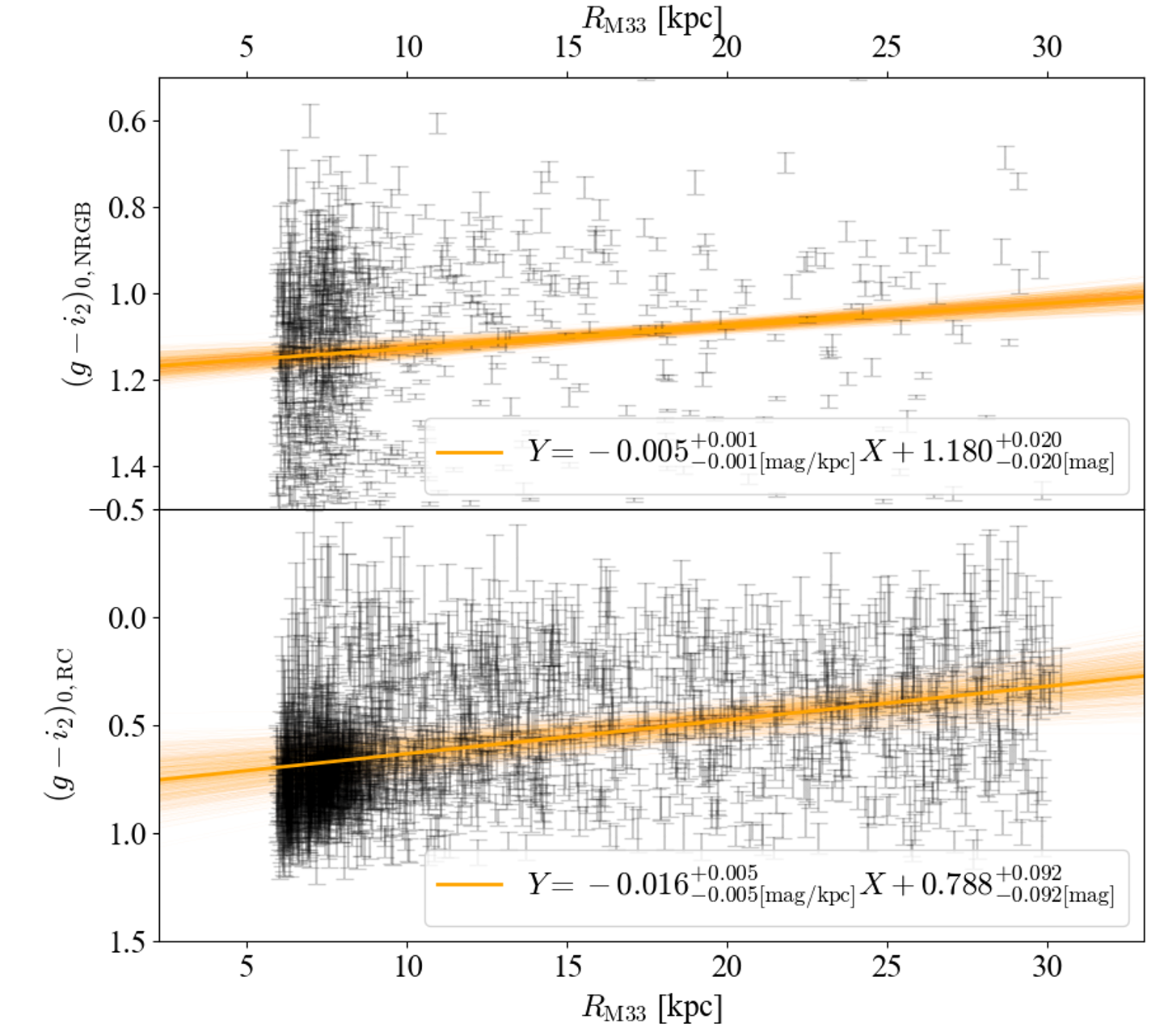}
\caption{The projected radial color profile. The top panel shows the color profile of the NRGB stars, and the bottom panel shows the color profile of the RC stars. The thick orange line shows the results of linear regression, and the thin orange lines indicate the results of 1,000 random samples from the posterior distribution.
\label{figure:ColorProfile}}
\end{figure*}

It has been known that the color of RGB/RC is sensitive to the age and metallicity of the stellar system. To confirm the differences in spatial gradient of the population in M33, we construct the color profile of NRGB/RC stars. Figure \ref{figure:ColorProfile} shows the constructed projected radial color profiles. To construct the color profile, we use only stars with $-0.25<\eta<0.25$ for simplicity as well as the radial density profile. The top panel indicates the color profile of the NRGB stars which is in the black polygon in Figure \ref{figure:CMD} (b), and the bottom panel shows the color profile of the RC stars which is in the red polygon in Figure \ref{figure:CMD} (c). In this figure, the black dot shows the color of each NRGB/RC sample and the black error bar shows its photometric error.

To examine the color gradient, we perform simple linear regression on this profile using the MCMC fitting. We initialize the sampler with 100 walkers, with 10,000 steps after a burn-in period of 100,000 steps to ensure a good sampling of the posterior distributions. We show the posterior and marginalized distributions in Appendix \ref{section:MCMC}. In Figure \ref{figure:ColorProfile}, we show the result using the median value of posterior distribution estimated by MCMC as the orange thick line. From this result, we confirm the existence of a color gradient in the NRGB/RC stars of M33, suggesting that the outer part of M33 has a more metal-poor and/or older population. In the Milky Way, an age gradient is found to exist in the disk \citep[e.g.,][]{martig2016}. However, our radial density profiles suggest that halo stars are more dominant than disk stars beyond $\sim 15$ kpc (see Figure \ref{figure:profile}), so the difference in color between the inner and outer regions suggests the presence of an old metal-poor population in the outer region of M33, which is possible to be the stellar halo.

\begin{figure*}[ht!]
\includegraphics[width=2\columnwidth]
{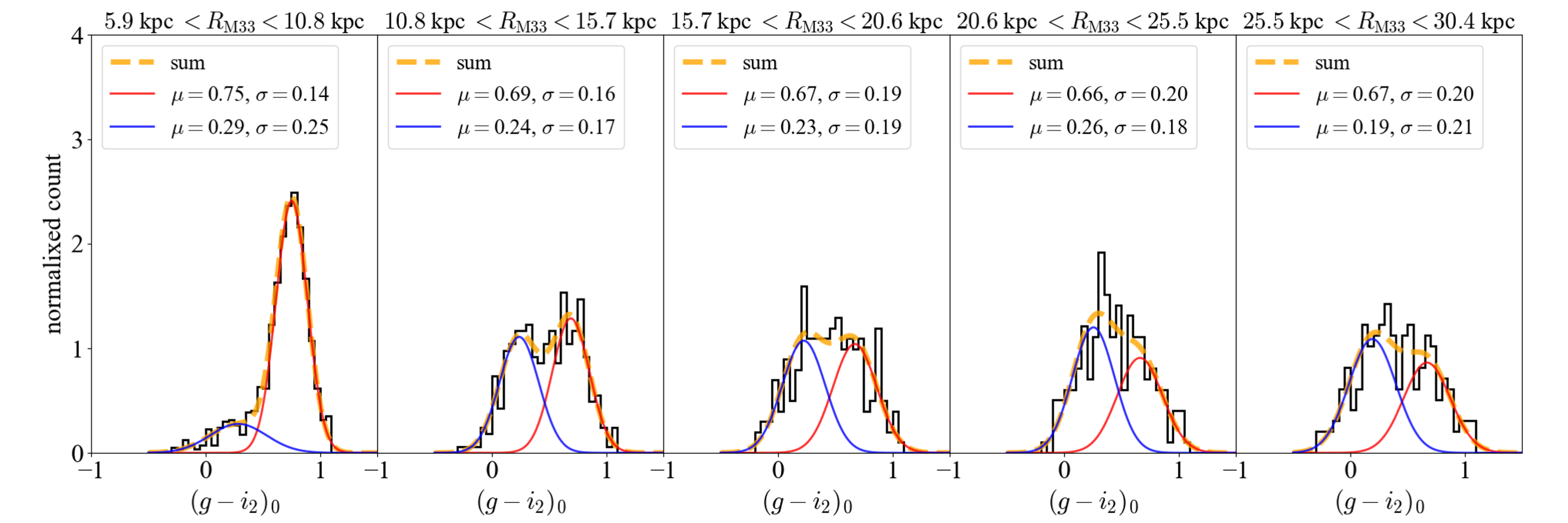}
\caption{The normalized color distributions of RC stars. The red and blue lines show the fitting results of two Gaussian distributions. The orange dashed line shows the summation of two Gaussian distributions.
\label{figure:ColorDistribution}}
\end{figure*}

\begin{figure}[ht!]
\includegraphics[width=1\columnwidth]
{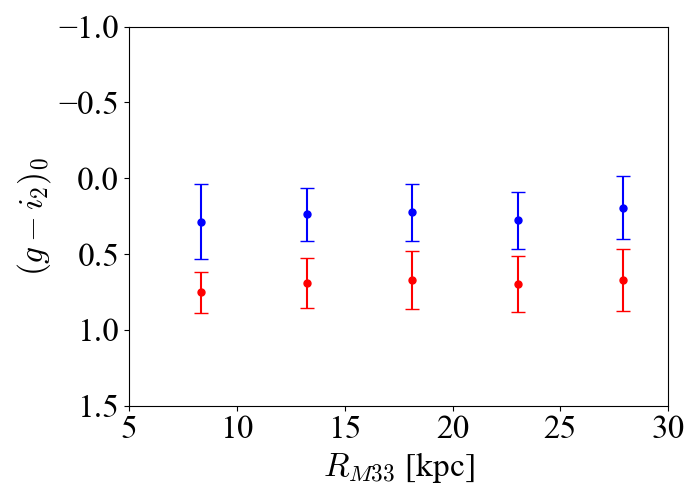}
\caption{The estimated mean and standard deviation of two Gaussian distributions from our RC stars as a function of distance from the center of M33. The color of these plots corresponds to the color of the lines in Figure \ref{figure:ColorDistribution}.
\label{figure:ColorDistribution2}}
\end{figure}

In order to investigate this color gradient, we construct normalized color distributions of the RC stars for each region as shown in Figure \ref{figure:ColorDistribution}. Figure \ref{figure:ColorDistribution} shows the normalized color distributions of the RC stars. The divided regions to construct these distributions are regions that roughly divide the HSC's field of view into five equal areas. In this color distribution, we can visually identify two components at $(g-i_2)_0 \sim 0.2$ and $(g-i_2)_0 \sim 0.7$. To confirm these components, we fit a double Gaussian model to these distributions. In Figure \ref{figure:ColorDistribution}, we show the fitting results as the red and blue lines. In Figure \ref{figure:ColorDistribution2}, we also show the estimated mean and standard deviation of two components as a function of distance from the center of M33. The color of plots in Figure \ref{figure:ColorDistribution2} corresponds to the color of the lines in Figure \ref{figure:ColorDistribution}. From these figures, we can confirm that two Gaussian distributions well reproduce our color distributions, and each population does not change from region to region. Also, it seems that the inner region is contaminated by red stars (i.e., young and/or metal-rich stars) and the outer region is contaminated by blue stars (i.e., old and/or metal-poor stars). Therefore, the color gradient shown in Figure \ref{figure:ColorProfile} is due to the fact that red stars, which are dominant in the inner region, sharply decrease toward the outer region, and blue stars are dominant in the outer area. Therefore, we think that, in the radial density profile of the RC stars, the extended population which is represented by a power-law consists of the blue stars (old and/or metal-poor stars). 

\section{Discussion}\label{section:Discussion}
\subsection{Shallow Stellar Halo}\label{section:OuterHalo}
As described in Section \ref{section:profile}, M33 has an extended stellar component like a stellar halo, and some radial profiles show shallow with $\alpha > -3$. Some of our fitted results show a shallow power-law, which differs from the result of \citet{smercina2023}. However, this is most likely because they focused on the inner region ($< 5$ kpc) and we focus on the outer region ($5 < R <30$ kpc), so we may see a different galactic structure.

In the Galactic halo, it is known that the halo is composed of double components: inner/outer halos or in-situ/ex-situ halos \citep[e.g.,][]{carollo2007,deason2014a,fukushima2019} which have different alpha indices. Also, the hydrodynamical simulation for galaxy formation by \citet{rodriguez-gomez2016} suggests that the simulated stellar halo consists of the in-situ and ex-situ halos with steep and shallow density slopes, respectively. It is worth noting from their simulation results that the transition radius for these halo components, $r_{\rm trans}$, is about five times a half-light radius for the stellar mass of $10^9~M_{\odot}$ (see their Figure 12), i.e., in our case with M33, we obtain $r_{\rm trans} = (5-7) \times R_d \sim 10-13~{\rm kpc}$. Thus, it is possible that the M33 halo has an inner halo at $R < r_{\rm trans}$ with a steep profile of $\alpha \sim -3$ and an outer halo at $R > r_{\rm trans}$ with a shallow slope of $\alpha > -3$ (Figure \ref{figure:profile} and \ref{figure:profile_434}). In this study, we also perform fitting with the double power law such as the broken power law, but MCMC does not converge well because there are too many free parameters. To settle this issue, it is expected that future spectroscopic observations \citep[e.g. Subaru/Prime Focus Spectrograph (PFS);][]{takada2014} of the outer regions will confirm whether the halo of M33 is well reproduced by the single halo or double halos. 

Our result shows that M33 has a shallow stellar halo in the outer region. However, for the Milky Way, the outer region up to $\sim 150$ kpc was estimated to have $\alpha = -3.74$ for single power-law or $-2.92$ for broken power-law \citep{fukushima2019}. Moreover, for M31, the halo profile was estimated to be ranging from $-3$ to $-2$ \citep{ibata2014}. Besides the large spirals in the Local Group, the HST survey also reported that nearby galaxies with stellar halo masses comparable to the Milky Way and M31 (NGC253, NGC891, NGC3031, NGC4565, NGC4945, and NGC 7814) have halo power-law slopes ranging from $-5$ to $-3$ \citep{harmsen2017}. \citet{pillepich2014a} found a strong correlation between stellar halo slope and total halo mass using the Illustris simulations. They predicted that more massive galaxies have shallower stellar haloes than the less massive ones. In this simulation result, they suggested that less massive galaxies with halo masses of $\sim 10^{11} M_{\odot}$ such as M33 have stellar halos reproduced by the power-law of $\alpha \sim -5$ (see their Figure 2). However, our results show that M33 has a shallow power-law population, so it suggests that M33 may have experienced a unique galactic formation mechanism if this prediction is correct.

One possible unique scenario of M33 to have a shallower population is the M31-M33 interaction. It has been suggested that M33 has interacted with M31 during the past 2-3 Gyr \citep{mcconnachie2009}. At this time, the M33 disk is thought to be extended by tidal forces. If such tidally-induced stretching occurred in the disk, the stellar halo may have been extended to form a shallow halo. Considering that a recent study has proposed that M33 is currently in the first infall \citep{patel2017}, it is necessary to verify the extent to which the halo is extended and to use orbital simulations to probe the extent to which dynamical interactions can stretch the stellar halo to flatten the inferred profile. We emphasize that we need spectroscopic observations (e.g., Subaru/PFS) in the future to verify whether the shallow profile identified in this study is due to a substructure hidden in M33. Besides this, it also needs comparison with other intermediate-mass galaxies (e.g., Large Magellanic Cloud and Small Magellanic Cloud) to verify whether the shallow profile is unique to intermediate-mass galaxies or common to spiral galaxies, including large galaxies. Finally, it should be noted that 

\subsection{Halo Properties}\label{section:HaloProperties}
Using the radial density profiles in Section \ref{section:DensityProfile}, we estimate the surface brightness of the M33 stellar halo by counting the flux of halo stars for regions up to $\sim 2.2$ deg ($\sim 30$ kpc). Due to the large dispersion of the contamination component for the RC stars (see Figure \ref{figure:SSP}), and for comparison with the large survey observations of M33, which observed only bright RGB \citep[PAndAS;][]{mcmonigal2016}, we limit to use the NRGB stars, which is selected based on {\it NB515} information and color-magnitude information (see Section \ref{section:RGB_RC}), for the derivation of the surface brightness of the M33 halo. If we include the RC stars to derive the surface brightness, the derived value would not change significantly, because the magnitude of our RC samples is $\sim$ 4 mag deeper (i.e. $\sim 30$ times fainter in the units of flux) than the TRGB (see, Figure \ref{figure:CMD}). To calculate the surface brightness, we construct the halo fraction profile as a function of distance from the M33 center using the estimated radial density profiles. Then, we count the flux of NRGB stars in each region weighted by the halo fraction. Following this method, we derive the surface brightness for each band. Finally, to compare with the PAndAS study, we convert to the {\it V}-band surface brightness using the transformation formula \citep[see Equation (8) in][]{komiyama2018}. The resulting surface brightness is $\mu_{\it V} = 35.72 \pm 0.08$ mag arcsec$^{-2}$. This value is consistent with the upper limit of $35.5$ for halo surface brightness suggested in \citet{mcmonigal2016}. Therefore, it is considered that the robust selection with the narrow-band filter allows us to detect the halo structure in this study. The global halo properties will be estimated using new data which is obtained from ongoing observation (Ogami et al. in prep).

\section{Conclusions}\label{section:Conclusion}
We have carried out deep multi-color imaging of the outskirts of M33 using Subaru/HSC. This observation covers a $\sim 1.76$ degree$^2$ field up to $\sim 30$ kpc from the center of M33, using {\it g-}, {\it $r_2$-}, {\it $i_2$-}bands, and {\it NB515}. We have extracted the point sources using the \texttt{extendedness} in the hscPipe and color-color diagram. For RC samples, we have conducted the color and magnitude selection. For RGB samples, we have calculated the probability of being an RGB star, in addition to the color and magnitude selection. Using these selections, we have succeeded in confirming the clear RGB sequence and RC in the CMD.

To confirm the properties of the outskirt in M33, we have constructed two types of radial profile: radial density profile for {\it NB515}-selected RGB (NRGB) stars and RC stars, and radial color profile for RC stars. We have performed the model fitting to the radial density profiles, and found that M33 has a shallower power-law profile with $\alpha > -3$ than large spiral galaxies and the previous study of M33 in some cases. Based on these results, we can present the following suggestions for the M31 stellar halo. First, regarding the halo profile of M33 being shallow compared to large spirals, it is possible that the M33 halo was influenced by the interaction with M31. As for reconciling our result with previous studies of M33, M33 may have a dual stellar halo structure and we might have detected the outer halo of M33 for the first time. In addition, the radial color profile shows that the color of RC stars is bluer toward the outer regions. This result suggests that the old and/or metal-poor population is more dominant in the outer regions, supporting that the stellar halo exists beyond 15 kpc.

Finally, we have derived the surface brightness of the stellar halo to compare with the previous study \citep{mcmonigal2016}. The surface brightness is $\mu_{\it V} = 35.72 \pm 0.08$, and this result is consistent with previous studies. We think that this is caused by the differences in sample selection, and this result shows that our narrow-band selection is a powerful tool to probe the low-surface brightness structure.

\begin{acknowledgments}
We acknowledge support in part from MEXT Grant-in-Aid for Scientific Research (No.~JP18H05437 and JP21H05448 for M.C., No.~JP21K13909 and JP23H04009 for K.H., and No.~JP22K14076 for T.K.). This work was partially supported by the Overseas Travel Fund for Students (2023) of the Astronomical Science Program, the Graduate University for Advanced Studies, SOKENDAI. Data analysis was in part carried out on the Large-scale data analysis system co-operated by the Astronomy Data Center (NAOJ/ADC) Subaru Telescope, NAOJ. E.N.K.\ acknowledges support from NSF CAREER grant AST-2233781. CF and RFGW are grateful for support through the generosity of Eric and Wendy Schmidt, by recommendation of the Schmidt Futures program.

\end{acknowledgments}

%

\vspace{5mm}
\facilities{Subaru (HSC)}

\software{emcee \citep{foreman-mackey2013},
          astropy \citep{theastropycollaboration2013},
          Matplotlib \citep{hunter2007},
          numpy \citep{vanderwalt2011},
          corner \citep{foreman-mackey2016}}


\newpage
\appendix
\section{MCMC Results}\label{section:MCMC}
\begin{figure}[ht!]
\centering
\includegraphics[width=0.91\columnwidth]
{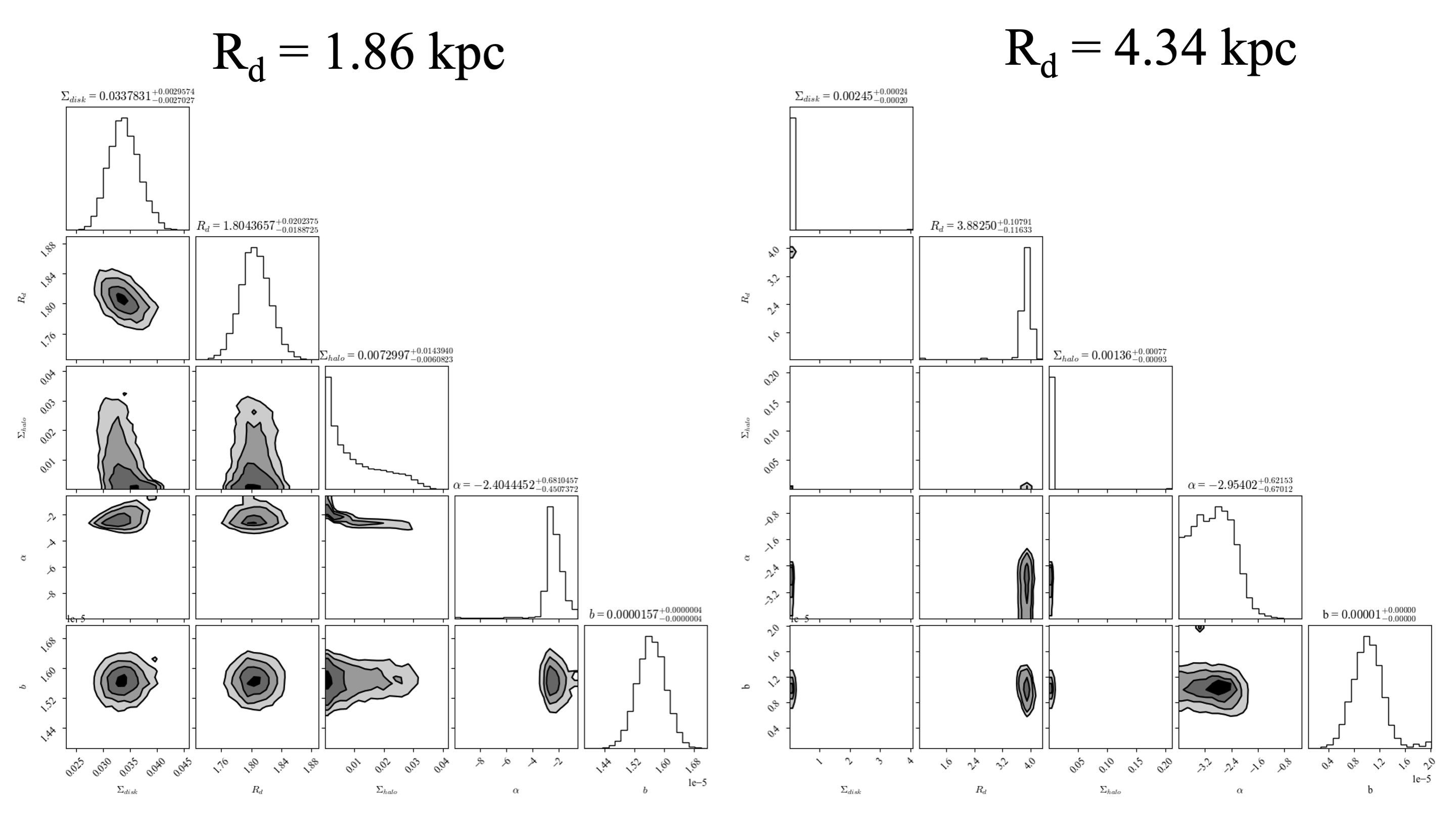}
\caption{Left: The posterior distributions and marginalized distributions in the radial density profile of NRGB stars, when we assume a prior distribution of $R_d$ as a Gaussian distribution of a mean with 1.86 kpc. Each panel shows the surface density scale of the disk, disk scale radius, surface density scale of the halo, projected power-law index, and contamination scale, from left (top) to right (bottom). Right: The same as in the left panel but an assumed prior for $R_d$ is a Gaussian distribution with a mean of 4.34 kpc.
\label{figure:MCMC_NRGB}}
\end{figure}

\begin{figure}[ht!]
\centering
\includegraphics[width=0.91\columnwidth]
{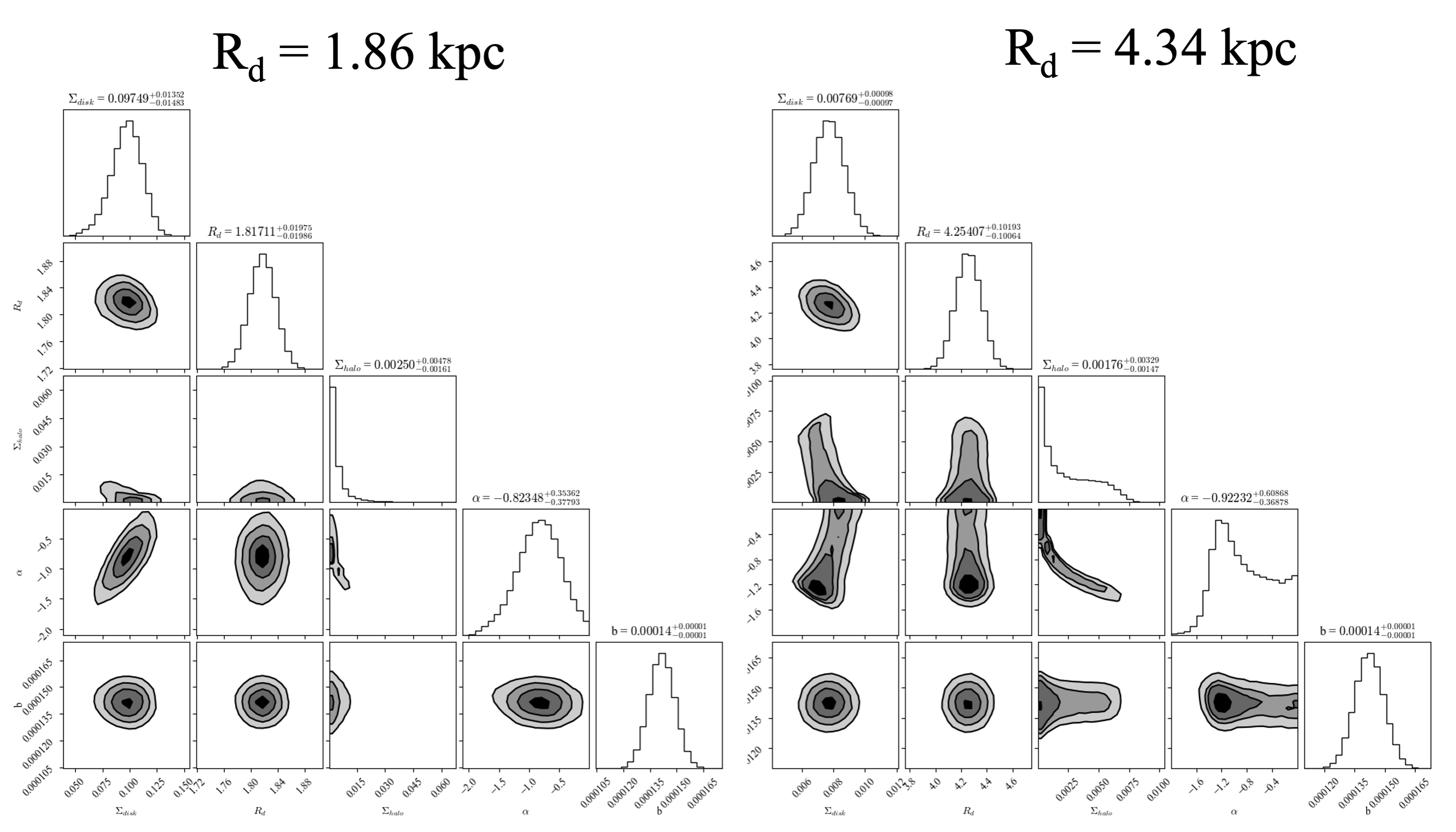}
\caption{The same as in Figure \ref{figure:MCMC_NRGB} but for the corner plot from MCMC fitting for the radial density profile of RC using the Gaussian prior with parameters derived from HSC-SSP data.
\label{figure:MCMC_RC}}
\end{figure}

\begin{figure}[ht!]
\centering
\includegraphics[width=0.96\columnwidth]
{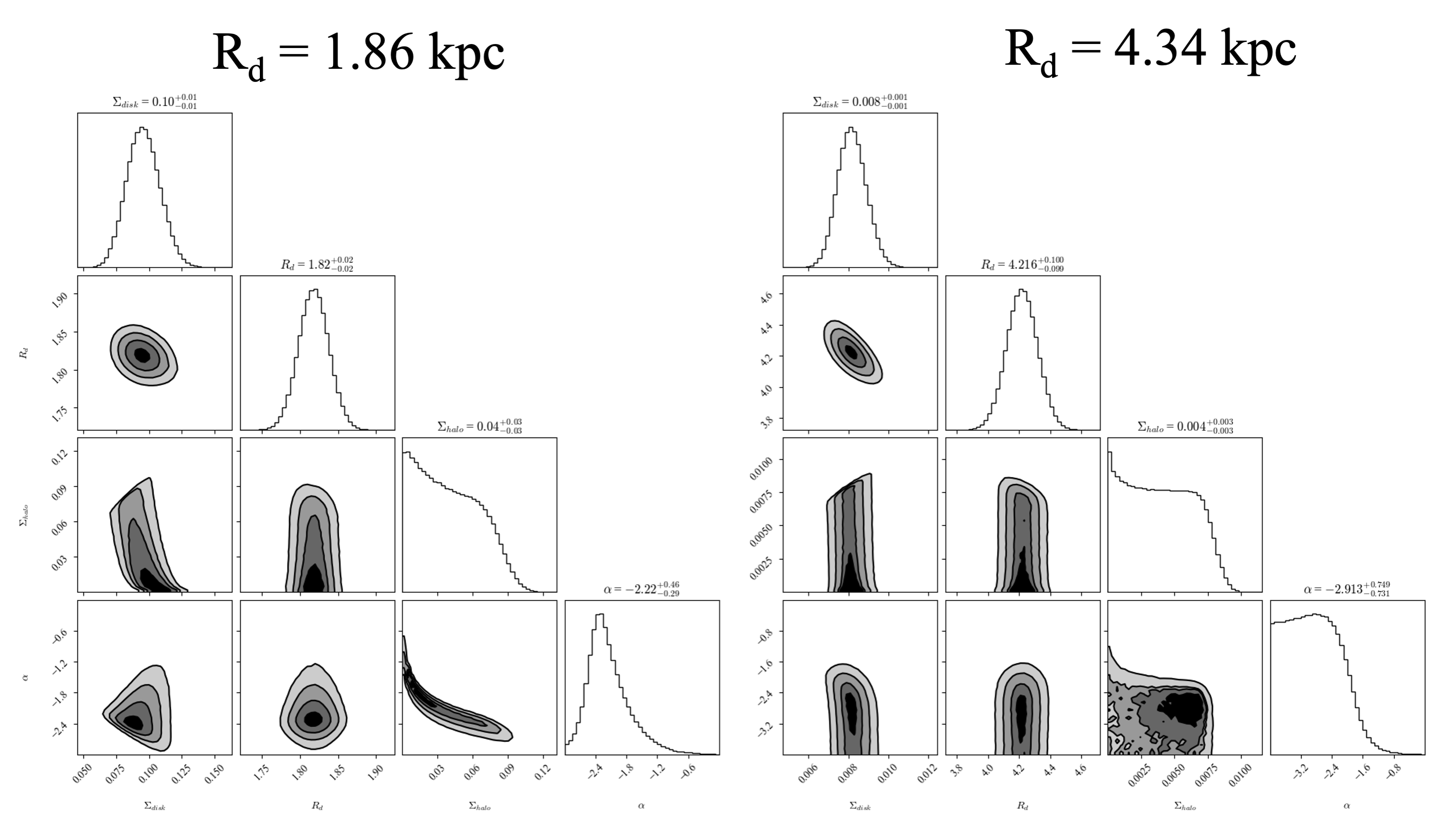}
\caption{The same as in Figure \ref{figure:MCMC_NRGB} but for the corner plot from MCMC fitting for the radial density profile of RC using the maximum value of the remaining background galaxies derived from HSC-SSP data.
\label{figure:MCMC_RC_SSPmax}}
\end{figure}

\begin{figure}[ht!]
\centering
\includegraphics[width=0.95\columnwidth]
{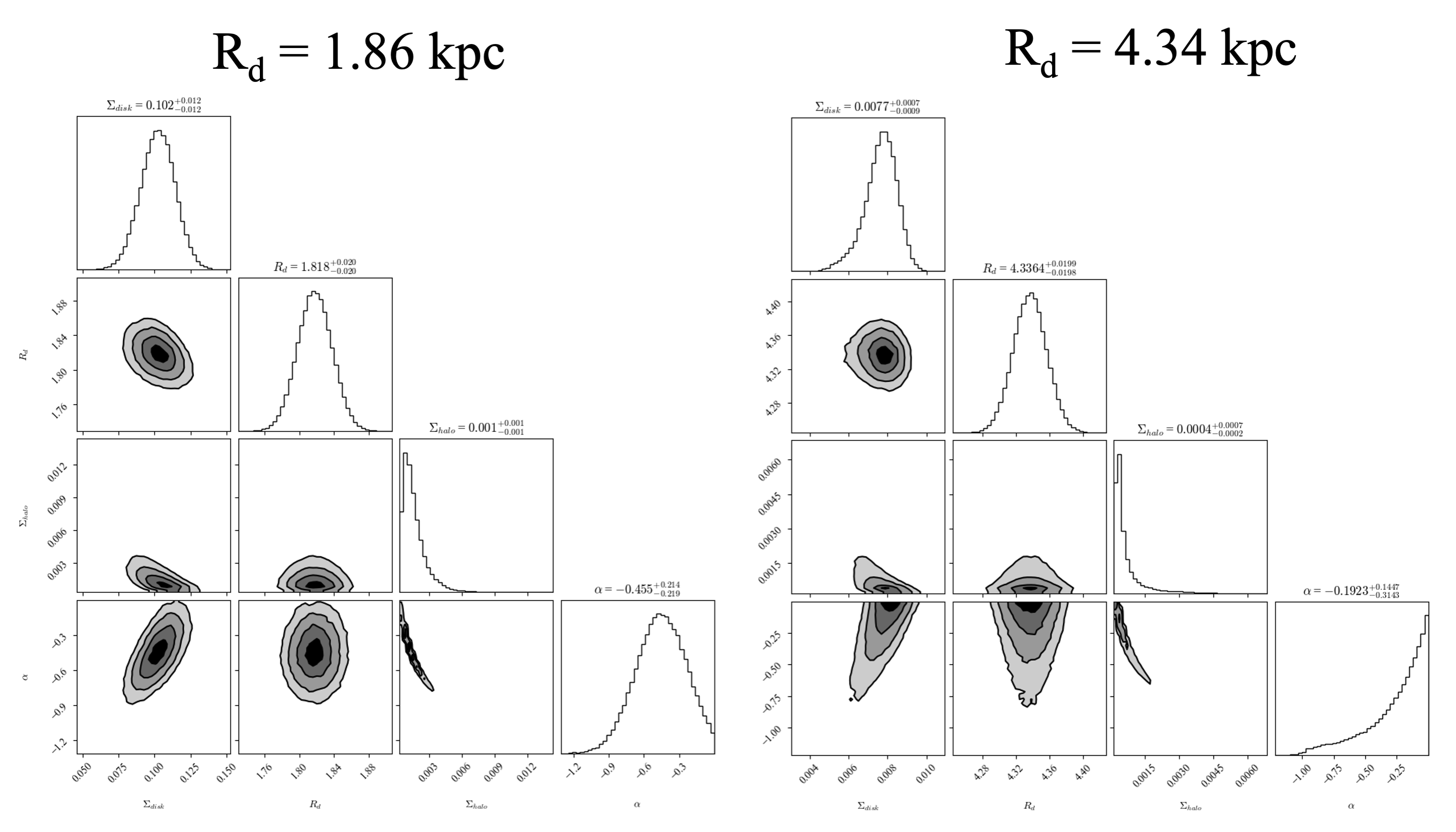}
\caption{The same as in Figure \ref{figure:MCMC_NRGB} but for the corner plot from MCMC fitting for the radial density profile of RC using the minimum value of the remaining background galaxies derived from HSC-SSP data.
\label{figure:MCMC_RC_SSPmin}}
\end{figure}

\begin{figure}[ht!]
\centering
\includegraphics[width=0.95\columnwidth]
{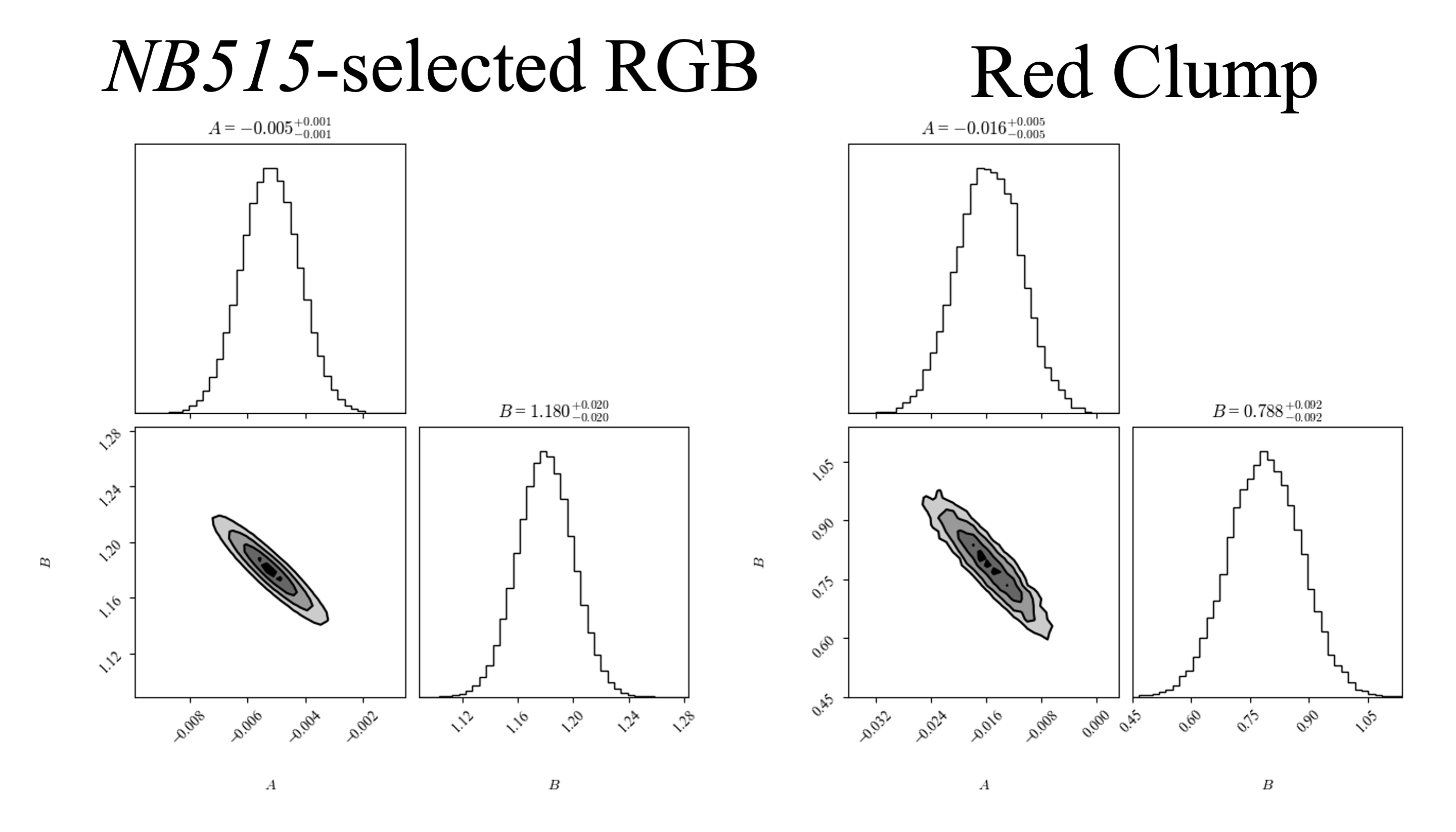}
\caption{Left: The posterior distributions and marginalized distributions from MCMC fitting for the radial color profile. Each panel shows the color slope of NRGB stars and the central color scale, from left (top) to right (bottom). Right: The same as in the left panel but for corner plot from MCMC fitting for the radial color profile of RC stars.
\label{figure:MCMC_ColorProfile}}
\end{figure}

Here, we show the MCMC results as corner plots (posterior distribution and marginal distributions for all parameters ) in the case of the radial density profiles (from Figure \ref{figure:MCMC_NRGB} to Figure \ref{figure:MCMC_RC_SSPmin}) and the radial color profile (Figure \ref{figure:MCMC_ColorProfile}).

\newpage
\bibliography{M33}{}

\begin{thebibliography}{}
\expandafter\ifx\csname natexlab\endcsname\relax\def\natexlab#1{#1}\fi
\providecommand{\url}[1]{\href{#1}{#1}}
\providecommand{\dodoi}[1]{doi:~\href{http://doi.org/#1}{\nolinkurl{#1}}}
\providecommand{\doeprint}[1]{\href{http://ascl.net/#1}{\nolinkurl{http://ascl.net/#1}}}
\providecommand{\doarXiv}[1]{\href{https://arxiv.org/abs/#1}{\nolinkurl{https://arxiv.org/abs/#1}}}

\bibitem[{Abbott {et~al.}(2018)Abbott, Abdalla, Allam, Amara, Annis, Asorey, Avila, Ballester, Banerji, Barkhouse, Baruah, Baumer, Bechtol, Becker, {Benoit-L{\'e}vy}, Bernstein, Bertin, Blazek, Bocquet, Brooks, Brout, {Buckley-Geer}, Burke, Busti, Campisano, {Cardiel-Sas}, Rosell, Kind, Carretero, Castander, Cawthon, Chang, Chen, Conselice, Costa, Crocce, Cunha, D'Andrea, da~Costa, Das, Daues, Davis, Davis, Vicente, DePoy, DeRose, Desai, Diehl, Dietrich, Dodelson, Doel, {Drlica-Wagner}, Eifler, Elliott, Evrard, Farahi, Neto, Fernandez, Finley, Flaugher, Foley, Fosalba, Friedel, Frieman, {Garc{\'i}a-Bellido}, Gaztanaga, Gerdes, Giannantonio, Gill, Glazebrook, Goldstein, Gower, Gruen, Gruendl, Gschwend, Gupta, Gutierrez, Hamilton, Hartley, Hinton, Hislop, Hollowood, Honscheid, Hoyle, Huterer, Jain, James, Jeltema, Johnson, Johnson, Kacprzak, Kent, Khullar, Klein, Kovacs, Koziol, Krause, Kremin, Kron, Kuehn, Kuhlmann, Kuropatkin, Lahav, Lasker, Li, Li, Liddle, Lima, Lin, {L{\'o}pez-Reyes}, MacCrann, Maia,
  Maloney, Manera, March, Marriner, Marshall, Martini, McClintock, McKay, McMahon, Melchior, Menanteau, Miller, Miquel, Mohr, Morganson, Mould, Neilsen, Nichol, Nogueira, Nord, Nugent, Nunes, Ogando, Old, Pace, Palmese, {Paz-Chinch{\'o}n}, Peiris, Percival, Petravick, Plazas, Poh, Pond, Porredon, Pujol, Refregier, Reil, Ricker, Rollins, Romer, Roodman, Rooney, Ross, Rykoff, Sako, Sanchez, Sanchez, Santiago, Saro, Scarpine, Scolnic, Serrano, {Sevilla-Noarbe}, Sheldon, Shipp, Silveira, Smith, Smith, Smith, {Soares-Santos}, Sobreira, Song, Stebbins, Suchyta, Sullivan, Swanson, Tarle, Thaler, Thomas, Thomas, Troxel, Tucker, Vikram, Vivas, Walker, Wechsler, Weller, Wester, Wolf, Wu, Yanny, Zenteno, Zhang, Zuntz, {DES Collaboration}, Juneau, Fitzpatrick, Nikutta, Nidever, Olsen, Scott, \& {NOAO Data Lab}}]{abbott2018}
Abbott, T. M.~C., Abdalla, F.~B., Allam, S., {et~al.} 2018, The Astrophysical Journal Supplement Series, 239, 18, \dodoi{10.3847/1538-4365/aae9f0}

\bibitem[{Ahumada(2020)}]{ahumada2020}
Ahumada, R. 2020, The Astrophysical Journal Supplement Series

\bibitem[{Aihara {et~al.}(2018{\natexlab{a}})Aihara, Armstrong, Bickerton, Bosch, Coupon, Furusawa, Hayashi, Ikeda, Kamata, Karoji, Kawanomoto, Koike, Komiyama, Lang, Lupton, Mineo, Miyatake, Miyazaki, Morokuma, Obuchi, Oishi, Okura, Price, Takata, Tanaka, Tanaka, Tanaka, Uchida, Uraguchi, Utsumi, Wang, Yamada, Yamanoi, Yasuda, Arimoto, Chiba, Finet, Fujimori, Fujimoto, Furusawa, Goto, Goulding, Gunn, Harikane, Hattori, Hayashi, He{\l}miniak, Higuchi, Hikage, Ho, Hsieh, Huang, Huang, Imanishi, Iwata, Jaelani, Jian, Kashikawa, Katayama, Kojima, Konno, Koshida, Kusakabe, Leauthaud, Lee, Lin, Lin, Mandelbaum, Matsuoka, Medezinski, Miyama, Momose, More, More, Mukae, Murata, Murayama, Nagao, Nakata, Niida, Niikura, Nishizawa, Oguri, Okabe, Ono, Onodera, Onoue, Ouchi, Pyo, Shibuya, Shimasaku, Simet, Speagle, Spergel, Strauss, Sugahara, Sugiyama, Suto, Suzuki, Tait, Takada, Terai, Toba, Turner, Uchiyama, Umetsu, Urata, Usuda, Yeh, \& Yuma}]{aihara2018a}
Aihara, H., Armstrong, R., Bickerton, S., {et~al.} 2018{\natexlab{a}}, Publications of the Astronomical Society of Japan, 70, \dodoi{10.1093/pasj/psx081}

\bibitem[{Aihara {et~al.}(2018{\natexlab{b}})Aihara, Arimoto, Armstrong, Arnouts, Bahcall, Bickerton, Bosch, Bundy, Capak, Chan, Chiba, Coupon, Egami, Enoki, Finet, Fujimori, Fujimoto, Furusawa, Furusawa, Goto, Goulding, Greco, Greene, Gunn, Hamana, Harikane, Hashimoto, Hattori, Hayashi, Hayashi, He{\l}miniak, Higuchi, Hikage, Ho, Hsieh, Huang, Huang, Ikeda, Imanishi, Inoue, Iwasawa, Iwata, Jaelani, Jian, Kamata, Karoji, Kashikawa, Katayama, Kawanomoto, Kayo, Koda, Koike, Kojima, Komiyama, Konno, Koshida, Koyama, Kusakabe, Leauthaud, Lee, Lin, Lin, Lupton, Mandelbaum, Matsuoka, Medezinski, Mineo, Miyama, Miyatake, Miyazaki, Momose, More, More, Moritani, Moriya, Morokuma, Mukae, Murata, Murayama, Nagao, Nakata, Niida, Niikura, Nishizawa, Obuchi, Oguri, Oishi, Okabe, Okura, Ono, Onodera, Onoue, Osato, Ouchi, Price, Pyo, Sako, Okamoto, Sawicki, Shibuya, Shimasaku, Shimono, Shirasaki, Silverman, Simet, Speagle, Spergel, Strauss, Sugahara, Sugiyama, Suto, Suyu, Suzuki, Tait, Takata, Takada, Tamura, Tanaka, Tanaka,
  Tanaka, Tanaka, Terai, Terashima, Toba, Toshikawa, Turner, Uchida, Uchiyama, Umetsu, Uraguchi, Urata, Usuda, Utsumi, Wang, Wang, Wong, Yabe, Yamada, Yamanoi, Yasuda, Yeh, Yonehara, \& Yuma}]{aihara2018}
Aihara, H., Arimoto, N., Armstrong, R., {et~al.} 2018{\natexlab{b}}, Publications of the Astronomical Society of Japan, 70, \dodoi{10.1093/pasj/psx066}

\bibitem[{Behroozi {et~al.}(2019)Behroozi, Wechsler, Hearin, \& Conroy}]{behroozi2019}
Behroozi, P., Wechsler, R.~H., Hearin, A.~P., \& Conroy, C. 2019, Monthly Notices of the Royal Astronomical Society, 488, 3143, \dodoi{10.1093/mnras/stz1182}

\bibitem[{Bosch {et~al.}(2018)Bosch, Armstrong, Bickerton, Furusawa, Ikeda, Koike, Lupton, Mineo, Price, Takata, Tanaka, Yasuda, AlSayyad, Becker, Coulton, Coupon, Garmilla, Huang, Krughoff, Lang, Leauthaud, Lim, Lust, MacArthur, Mandelbaum, Miyatake, Miyazaki, Murata, More, Okura, Owen, Swinbank, Strauss, Yamada, \& Yamanoi}]{bosch2018}
Bosch, J., Armstrong, R., Bickerton, S., {et~al.} 2018, Publications of the Astronomical Society of Japan, 70, \dodoi{10.1093/pasj/psx080}

\bibitem[{Bressan {et~al.}(2012)Bressan, Marigo, Girardi, Salasnich, Dal~Cero, Rubele, \& Nanni}]{bressan2012}
Bressan, A., Marigo, P., Girardi, L., {et~al.} 2012, Monthly Notices of the Royal Astronomical Society, 427, 127, \dodoi{10.1111/j.1365-2966.2012.21948.x}

\bibitem[{Bullock \& Johnston(2005)}]{bullock2005}
Bullock, J.~S., \& Johnston, K.~V. 2005, The Astrophysical Journal, 635, 931, \dodoi{10.1086/497422}

\bibitem[{Carollo {et~al.}(2007)Carollo, Beers, Lee, Chiba, Norris, Wilhelm, Sivarani, Marsteller, Munn, {Bailer-Jones}, Fiorentin, \& York}]{carollo2007}
Carollo, D., Beers, T.~C., Lee, Y.~S., {et~al.} 2007, Nature, 450, 1020, \dodoi{10.1038/nature06460}

\bibitem[{Chambers {et~al.}(2016)Chambers, Magnier, Metcalfe, Flewelling, Huber, Waters, Denneau, Draper, Farrow, Finkbeiner, Holmberg, Koppenhoefer, Price, Rest, Saglia, Schlafly, Smartt, Sweeney, Wainscoat, Burgett, Chastel, Grav, Heasley, Hodapp, Jedicke, Kaiser, Kudritzki, Luppino, Lupton, Monet, Morgan, Onaka, Shiao, Stubbs, Tonry, White, Ba{\~n}ados, Bell, Bender, Bernard, Boegner, Boffi, Botticella, Calamida, Casertano, Chen, Chen, Cole, Deacon, Frenk, Fitzsimmons, Gezari, Gibbs, Goessl, Goggia, Gourgue, Goldman, Grant, Grebel, Hambly, Hasinger, Heavens, Heckman, Henderson, Henning, Holman, Hopp, Ip, Isani, Jackson, Keyes, Koekemoer, Kotak, Le, Liska, Long, Lucey, Liu, Martin, Masci, McLean, Mindel, Misra, Morganson, Murphy, Obaika, Narayan, {Nieto-Santisteban}, Norberg, Peacock, Pier, Postman, Primak, Rae, Rai, Riess, Riffeser, Rix, R{\"o}ser, Russel, Rutz, Schilbach, Schultz, Scolnic, Strolger, Szalay, Seitz, Small, Smith, Soderblom, Taylor, Thomson, Taylor, Thakar, Thiel, Thilker, Unger, Urata,
  Valenti, Wagner, Walder, Walter, Watters, Werner, {Wood-Vasey}, \& Wyse}]{chambers2016}
Chambers, K.~C., Magnier, E.~A., Metcalfe, N., {et~al.} 2016, The {{Pan-STARRS1 Surveys}}, \dodoi{10.48550/arXiv.1612.05560}

\bibitem[{Corbelli {et~al.}(2014)Corbelli, Thilker, Zibetti, Giovanardi, \& Salucci}]{corbelli2014}
Corbelli, E., Thilker, D., Zibetti, S., Giovanardi, C., \& Salucci, P. 2014, Astronomy \& Astrophysics, 572, A23, \dodoi{10.1051/0004-6361/201424033}

\bibitem[{{de~Grijs} {et~al.}(2017){de~Grijs}, Courbin, {Mart{\'i}nez-V{\'a}zquez}, Monelli, Oguri, \& Suyu}]{degrijs2017}
{de~Grijs}, R., Courbin, F., {Mart{\'i}nez-V{\'a}zquez}, C.~E., {et~al.} 2017, Space Science Reviews, 212, 1743, \dodoi{10.1007/s11214-017-0395-z}

\bibitem[{Deason {et~al.}(2014)Deason, Belokurov, Koposov, \& Rockosi}]{deason2014a}
Deason, A.~J., Belokurov, V., Koposov, S.~E., \& Rockosi, C.~M. 2014, The Astrophysical Journal

\bibitem[{Fitzpatrick(1999)}]{fitzpatrick1999}
Fitzpatrick, E.~L. 1999, Publications of the Astronomical Society of the Pacific, 111, 63, \dodoi{10.1086/316293}

\bibitem[{Flewelling {et~al.}(2020)Flewelling, Magnier, Chambers, Heasley, Holmberg, Huber, Sweeney, Waters, Calamida, Casertano, Chen, Farrow, Hasinger, Henderson, Long, Metcalfe, Narayan, {Nieto-Santisteban}, Norberg, Rest, Saglia, Szalay, Thakar, Tonry, Valenti, Werner, White, Denneau, Draper, Hodapp, Jedicke, Kaiser, Kudritzki, Price, Wainscoat, Chastel, McLean, Postman, \& Shiao}]{flewelling2020}
Flewelling, H.~A., Magnier, E.~A., Chambers, K.~C., {et~al.} 2020, The Astrophysical Journal Supplement Series, 251, 7, \dodoi{10.3847/1538-4365/abb82d}

\bibitem[{{Foreman-Mackey}(2016)}]{foreman-mackey2016}
{Foreman-Mackey}, D. 2016, The Journal of Open Source Software, 1, 24, \dodoi{10.21105/joss.00024}

\bibitem[{{Foreman-Mackey} {et~al.}(2013){Foreman-Mackey}, Hogg, Lang, \& Goodman}]{foreman-mackey2013}
{Foreman-Mackey}, D., Hogg, D.~W., Lang, D., \& Goodman, J. 2013, Publications of the Astronomical Society of the Pacific, 125, 306, \dodoi{10.1086/670067}

\bibitem[{Fukushima {et~al.}(2018)Fukushima, Chiba, Homma, Okamoto, Komiyama, Tanaka, Tanaka, Arimoto, \& Matsuno}]{fukushima2018}
Fukushima, T., Chiba, M., Homma, D., {et~al.} 2018, Publications of the Astronomical Society of Japan, 70, \dodoi{10.1093/pasj/psy060}

\bibitem[{Fukushima {et~al.}(2019)Fukushima, Chiba, Tanaka, Hayashi, Homma, Okamoto, Komiyama, Tanaka, Arimoto, \& Matsuno}]{fukushima2019}
Fukushima, T., Chiba, M., Tanaka, M., {et~al.} 2019, Publications of the Astronomical Society of Japan, 71, 72, \dodoi{10.1093/pasj/psz052}

\bibitem[{{Gaia Collaboration} {et~al.}(2016){Gaia Collaboration}, Brown, Vallenari, Prusti, De~Bruijne, Mignard, Drimmel, Babusiaux, {Bailer-Jones}, Bastian, Biermann, Evans, Eyer, Jansen, Jordi, Katz, Klioner, Lammers, Lindegren, Luri, O'Mullane, Panem, Pourbaix, Randich, Sartoretti, Siddiqui, Soubiran, Valette, Van~Leeuwen, Walton, Aerts, Arenou, Cropper, H{\o}g, Lattanzi, Grebel, Holland, Huc, Passot, Perryman, Bramante, Cacciari, Casta{\~n}eda, Chaoul, Cheek, De~Angeli, Fabricius, Guerra, Hern{\'a}ndez, {Jean-Antoine-Piccolo}, Masana, Messineo, Mowlavi, Nienartowicz, {Ord{\'o}{\~n}ez-Blanco}, Panuzzo, Portell, Richards, Riello, Seabroke, Tanga, Th{\'e}venin, Torra, Els, {Gracia-Abril}, Comoretto, {Garcia-Reinaldos}, Lock, Mercier, Altmann, Andrae, Astraatmadja, {Bellas-Velidis}, Benson, Berthier, Blomme, Busso, Carry, Cellino, Clementini, Cowell, Creevey, Cuypers, Davidson, De~Ridder, De~Torres, Delchambre, Dell'Oro, Ducourant, Fr{\'e}mat, {Garc{\'i}a-Torres}, Gosset, Halbwachs, Hambly, Harrison, Hauser,
  Hestroffer, Hodgkin, Huckle, Hutton, Jasniewicz, Jordan, Kontizas, Korn, Lanzafame, Manteiga, Moitinho, Muinonen, Osinde, Pancino, Pauwels, Petit, {Recio-Blanco}, Robin, Sarro, Siopis, Smith, Smith, Sozzetti, Thuillot, Van~Reeven, Viala, Abbas, Abreu~Aramburu, Accart, Aguado, Allan, Allasia, Altavilla, {\'A}lvarez, Alves, Anderson, Andrei, Anglada~Varela, Antiche, Antoja, Ant{\'o}n, Arcay, Bach, Baker, {Balaguer-N{\'u}{\~n}ez}, Barache, Barata, Barbier, Barblan, Barrado Y~Navascu{\'e}s, Barros, Barstow, Becciani, Bellazzini, Bello~Garc{\'i}a, Belokurov, Bendjoya, Berihuete, Bianchi, Bienaym{\'e}, Billebaud, Blagorodnova, {Blanco-Cuaresma}, Boch, Bombrun, Borrachero, Bouquillon, Bourda, Bouy, Bragaglia, Breddels, Brouillet, Br{\"u}semeister, Bucciarelli, Burgess, Burgon, Burlacu, Busonero, Buzzi, Caffau, Cambras, Campbell, Cancelliere, {Cantat-Gaudin}, Carlucci, Carrasco, Castellani, Charlot, Charnas, Chiavassa, Clotet, Cocozza, Collins, Costigan, Crifo, Cross, Crosta, Crowley, Dafonte, Damerdji, Dapergolas,
  David, David, De~Cat, De~Felice, De~Laverny, De~Luise, De~March, De~Martino, De~Souza, Debosscher, Del~Pozo, Delbo, Delgado, Delgado, Di~Matteo, Diakite, Distefano, Dolding, Dos~Anjos, Drazinos, Duran, Dzigan, Edvardsson, Enke, Evans, Eynard~Bontemps, Fabre, Fabrizio, Faigler, Falc{\~a}o, Farr{\`a}s~Casas, Federici, Fedorets, {Fern{\'a}ndez-Hern{\'a}ndez}, Fernique, Fienga, Figueras, Filippi, Findeisen, Fonti, Fouesneau, Fraile, Fraser, Fuchs, Gai, Galleti, Galluccio, Garabato, {Garc{\'i}a-Sedano}, Garofalo, Garralda, Gavras, Gerssen, Geyer, Gilmore, Girona, Giuffrida, Gomes, {Gonz{\'a}lez-Marcos}, {Gonz{\'a}lez-N{\'u}{\~n}ez}, {Gonz{\'a}lez-Vidal}, Granvik, Guerrier, Guillout, Guiraud, G{\'u}rpide, {Guti{\'e}rrez-S{\'a}nchez}, Guy, Haigron, Hatzidimitriou, Haywood, Heiter, Helmi, Hobbs, Hofmann, Holl, Holland, Hunt, Hypki, Icardi, Irwin, Jevardat De~Fombelle, Jofr{\'e}, Jonker, Jorissen, Julbe, Karampelas, Kochoska, Kohley, Kolenberg, Kontizas, Koposov, Kordopatis, Koubsky, {Krone-Martins}, Kudryashova,
  Kull, Bachchan, {Lacoste-Seris}, Lanza, Lavigne, {Le Poncin-Lafitte}, Lebreton, Lebzelter, Leccia, Leclerc, {Lecoeur-Taibi}, Lemaitre, Lenhardt, Leroux, Liao, Licata, Lindstr{\o}m, Lister, Livanou, Lobel, L{\"o}ffler, L{\'o}pez, Lorenz, MacDonald, Magalh{\~a}es~Fernandes, Managau, Mann, Mantelet, Marchal, Marchant, Marconi, Marinoni, Marrese, Marschalk{\'o}, Marshall, {Mart{\'i}n-Fleitas}, Martino, Mary, Matijevi{\v c}, Mazeh, McMillan, Messina, Michalik, Millar, Miranda, Molina, Molinaro, Molinaro, Moln{\'a}r, Moniez, Montegriffo, Mor, Mora, Morbidelli, Morel, Morgenthaler, Morris, Mulone, Muraveva, Musella, Narbonne, Nelemans, Nicastro, Noval, Ord{\'e}novic, {Ordieres-Mer{\'e}}, Osborne, Pagani, Pagano, Pailler, Palacin, Palaversa, Parsons, Pecoraro, Pedrosa, Pentik{\"a}inen, Pichon, Piersimoni, Pineau, Plachy, Plum, Poujoulet, Pr{\v s}a, Pulone, Ragaini, Rago, Rambaux, {Ramos-Lerate}, Ranalli, Rauw, Read, Regibo, Reyl{\'e}, Ribeiro, Rimoldini, Ripepi, Riva, Rixon, Roelens, {Romero-G{\'o}mez}, Rowell,
  Royer, {Ruiz-Dern}, Sadowski, Sagrist{\`a}~Sell{\'e}s, Sahlmann, Salgado, Salguero, Sarasso, Savietto, Schultheis, Sciacca, Segol, Segovia, Segransan, Shih, Smareglia, Smart, Solano, Solitro, Sordo, Soria~Nieto, Souchay, Spagna, Spoto, Stampa, Steele, Steidelm{\"u}ller, Stephenson, Stoev, Suess, S{\"u}veges, Surdej, Szabados, {Szegedi-Elek}, Tapiador, Taris, Tauran, Taylor, Teixeira, Terrett, Tingley, Trager, Turon, Ulla, Utrilla, Valentini, Van~Elteren, Van~Hemelryck, Van~Leeuwen, Varadi, Vecchiato, Veljanoski, Via, Vicente, Vogt, Voss, Votruba, Voutsinas, Walmsley, Weiler, Weingrill, Wevers, Wyrzykowski, Yoldas, {\v Z}erjal, Zucker, Zurbach, Zwitter, Alecu, Allen, Allende~Prieto, Amorim, {Anglada-Escud{\'e}}, Arsenijevic, Azaz, Balm, Beck, Bernstein, Bigot, Bijaoui, Blasco, Bonfigli, Bono, Boudreault, Bressan, Brown, Brunet, Bunclark, Buonanno, Butkevich, Carret, Carrion, Chemin, Ch{\'e}reau, Corcione, Darmigny, De~Boer, De~Teodoro, De~Zeeuw, Delle~Luche, Domingues, Dubath, Fodor, Fr{\'e}zouls, Fries,
  Fustes, Fyfe, Gallardo, Gallegos, Gardiol, Gebran, Gomboc, G{\'o}mez, Grux, Gueguen, Heyrovsky, Hoar, Iannicola, Isasi~Parache, Janotto, Joliet, Jonckheere, Keil, Kim, Klagyivik, Klar, Knude, Kochukhov, Kolka, Kos, Kutka, Lainey, LeBouquin, Liu, Loreggia, Makarov, Marseille, Martayan, {Martinez-Rubi}, Massart, Meynadier, Mignot, Munari, Nguyen, Nordlander, Ocvirk, O'Flaherty, Olias~Sanz, Ortiz, Osorio, Oszkiewicz, Ouzounis, Palmer, Park, Pasquato, Peltzer, Peralta, P{\'e}turaud, Pieniluoma, Pigozzi, Poels, Prat, Prod'homme, Raison, Rebordao, Risquez, {Rocca-Volmerange}, Rosen, {Ruiz-Fuertes}, Russo, Sembay, Serraller~Vizcaino, Short, Siebert, Silva, Sinachopoulos, Slezak, Soffel, Sosnowska, Strai{\v z}ys, Ter~Linden, Terrell, Theil, Tiede, Troisi, Tsalmantza, Tur, Vaccari, Vachier, Valles, Van~Hamme, Veltz, Virtanen, Wallut, Wichmann, Wilkinson, Ziaeepour, \& Zschocke}]{gaiacollaboration2016}
{Gaia Collaboration}, Brown, A. G.~A., Vallenari, A., {et~al.} 2016, Astronomy \& Astrophysics, 595, A2, \dodoi{10.1051/0004-6361/201629512}

\bibitem[{{Galera-Rosillo} {et~al.}(2018){Galera-Rosillo}, Corradi, \& Mampaso}]{galera-rosillo2018}
{Galera-Rosillo}, R., Corradi, R. L.~M., \& Mampaso, A. 2018, Astronomy \& Astrophysics, 612, A35, \dodoi{10.1051/0004-6361/201731383}

\bibitem[{Gilbert {et~al.}(2009)Gilbert, Guhathakurta, Kollipara, Beaton, Geha, Kalirai, Kirby, Majewski, \& Patterson}]{gilbert2009}
Gilbert, K.~M., Guhathakurta, P., Kollipara, P., {et~al.} 2009, The Astrophysical Journal, 705, 1275, \dodoi{10.1088/0004-637X/705/2/1275}

\bibitem[{Gilbert {et~al.}(2012)Gilbert, Guhathakurta, Beaton, Bullock, Geha, Kalirai, Kirby, Majewski, Ostheimer, Patterson, Tollerud, Tanaka, \& Chiba}]{gilbert2012}
Gilbert, K.~M., Guhathakurta, P., Beaton, R.~L., {et~al.} 2012, The Astrophysical Journal, 760, 76, \dodoi{10.1088/0004-637X/760/1/76}

\bibitem[{Gilbert {et~al.}(2022)Gilbert, Quirk, Guhathakurta, Tollerud, Wojno, Dalcanton, Durbin, Seth, Williams, Fung, Tangirala, \& Yusufali}]{gilbert2022}
Gilbert, K.~M., Quirk, A. C.~N., Guhathakurta, P., {et~al.} 2022, The Astrophysical Journal, 924, 116, \dodoi{10.3847/1538-4357/ac3480}

\bibitem[{Harmsen {et~al.}(2017)Harmsen, Monachesi, Bell, De~Jong, Bailin, {Radburn-Smith}, \& Holwerda}]{harmsen2017}
Harmsen, B., Monachesi, A., Bell, E.~F., {et~al.} 2017, Monthly Notices of the Royal Astronomical Society, 466, 1491, \dodoi{10.1093/mnras/stw2992}

\bibitem[{Homma {et~al.}(2016)Homma, Chiba, Okamoto, Komiyama, Tanaka, Tanaka, Ishigaki, Akiyama, Arimoto, Garmilla, Lupton, Strauss, Furusawa, Miyazaki, Murayama, Nishizawa, Takada, Usuda, \& Wang}]{homma2016}
Homma, D., Chiba, M., Okamoto, S., {et~al.} 2016, The Astrophysical Journal, 832, 21, \dodoi{10.3847/0004-637X/832/1/21}

\bibitem[{Homma {et~al.}(2023)Homma, Chiba, Komiyama, Tanaka, Okamoto, Tanaka, Ishigaki, Hayashi, Arimoto, Lupton, Strauss, Miyazaki, Wang, \& Murayama}]{homma2023}
Homma, D., Chiba, M., Komiyama, Y., {et~al.} 2023, Final {{Results}} of {{Search}} for {{New Milky Way Satellites}} in the {{Hyper Suprime-Cam Subaru Strategic Program Survey}}: {{Discovery}} of {{Two More Candidates}},  arXiv, \dodoi{10.48550/arXiv.2311.05439}

\bibitem[{Hunter(2007)}]{hunter2007}
Hunter, J.~D. 2007, Computing in Science \& Engineering, 9, 90, \dodoi{10.1109/MCSE.2007.55}

\bibitem[{Ibata {et~al.}(2007)Ibata, Martin, Irwin, Chapman, Ferguson, Lewis, \& McConnachie}]{ibata2007}
Ibata, R., Martin, N.~F., Irwin, M., {et~al.} 2007, The Astrophysical Journal, 671, 1591, \dodoi{10.1086/522574}

\bibitem[{Ibata {et~al.}(2014)Ibata, Lewis, McConnachie, Martin, Irwin, Ferguson, Babul, Bernard, Chapman, Collins, Fardal, Mackey, Navarro, Pe{\~n}arrubia, Rich, Tanvir, \& Widrow}]{ibata2014}
Ibata, R.~A., Lewis, G.~F., McConnachie, A.~W., {et~al.} 2014, The Astrophysical Journal, 780, 128, \dodoi{10.1088/0004-637X/780/2/128}

\bibitem[{Ivezic {et~al.}(2008)Ivezic, Axelrod, Brandt, Burke, Claver, Connolly, Cook, Gee, Gilmore, Jacoby, Jones, Kahn, Kantor, Krabbendam, Lupton, Monet, Pinto, Saha, Schalk, \& Schneider}]{ivezic2008}
Ivezic, Z., Axelrod, T., Brandt, W., {et~al.} 2008, Serbian Astronomical Journal, 1, \dodoi{10.2298/SAJ0876001I}

\bibitem[{Jang {et~al.}(2020)Jang, De~Jong, Holwerda, Monachesi, Bell, \& Bailin}]{jang2020}
Jang, I.~S., De~Jong, R.~S., Holwerda, B.~W., {et~al.} 2020, Astronomy \& Astrophysics, 637, A8, \dodoi{10.1051/0004-6361/201936994}

\bibitem[{Juri{\'c} {et~al.}(2008)Juri{\'c}, Ivezi{\'c}, Brooks, Lupton, Schlegel, Finkbeiner, Padmanabhan, Bond, Sesar, Rockosi, Knapp, Gunn, Sumi, Schneider, Barentine, Brewington, Brinkmann, Fukugita, Harvanek, Kleinman, Krzesinski, Long, Neilsen, Nitta, Snedden, \& York}]{juric2008}
Juri{\'c}, M., Ivezi{\'c}, {\v Z}., Brooks, A., {et~al.} 2008, The Astrophysical Journal, 673, 864, \dodoi{10.1086/523619}

\bibitem[{Kam {et~al.}(2015)Kam, Carignan, Chemin, Amram, \& Epinat}]{kam2015}
Kam, Z.~S., Carignan, C., Chemin, L., Amram, P., \& Epinat, B. 2015, Monthly Notices of the Royal Astronomical Society, 449, 4048, \dodoi{10.1093/mnras/stv517}

\bibitem[{Komiyama {et~al.}(2018)Komiyama, Chiba, Tanaka, Tanaka, Kirihara, Miki, Mori, Lupton, Guhathakurta, Kalirai, Gilbert, Kirby, Lee, Jang, Sharma, \& Hayashi}]{komiyama2018}
Komiyama, Y., Chiba, M., Tanaka, M., {et~al.} 2018, The Astrophysical Journal, 853, 29, \dodoi{10.3847/1538-4357/aaa129}

\bibitem[{Krisciunas {et~al.}(1998)Krisciunas, Margon, \& Szkody}]{krisciunas1998}
Krisciunas, K., Margon, B., \& Szkody, P. 1998, Publications of the Astronomical Society of the Pacific, 110, 1342, \dodoi{10.1086/316264}

\bibitem[{Lenz {et~al.}(1998)Lenz, Newberg, Rosner, Richards, \& Stoughton}]{lenz1998}
Lenz, D.~D., Newberg, J., Rosner, R., Richards, G.~T., \& Stoughton, C. 1998, The Astrophysical Journal Supplement Series, 119, 121, \dodoi{10.1086/313155}

\bibitem[{Lyke {et~al.}(2020)Lyke, Higley, McLane, Schurhammer, Myers, Ross, Dawson, Chabanier, Martini, Busca, Mas Des~Bourboux, Salvato, Streblyanska, Zarrouk, Burtin, Anderson, Bautista, Bizyaev, Brandt, Brinkmann, Brownstein, Comparat, Green, Macorra, Guti{\'e}rrez, Hou, Newman, {Palanque-Delabrouille}, P{\^a}ris, Percival, Petitjean, Rich, Rossi, Schneider, Smith, Vivek, \& Weaver}]{lyke2020}
Lyke, B.~W., Higley, A.~N., McLane, J.~N., {et~al.} 2020, The Astrophysical Journal Supplement Series, 250, 8, \dodoi{10.3847/1538-4365/aba623}

\bibitem[{Magnier {et~al.}(2013)Magnier, Schlafly, Finkbeiner, Juric, Tonry, Burgett, Chambers, Flewelling, Kaiser, Kudritzki, Morgan, Price, Sweeney, \& Stubbs}]{magnier2013}
Magnier, E.~A., Schlafly, E., Finkbeiner, D., {et~al.} 2013, The Astrophysical Journal Supplement Series, 205, 20, \dodoi{10.1088/0067-0049/205/2/20}

\bibitem[{Majewski {et~al.}(2000)Majewski, Ostheimer, Kunkel, \& Patterson}]{majewski2000}
Majewski, S.~R., Ostheimer, J.~C., Kunkel, W.~E., \& Patterson, R.~J. 2000, The Astronomical Journal, 120, 2550, \dodoi{10.1086/316836}

\bibitem[{Marigo(2017)}]{marigo2017}
Marigo, P. 2017, The Astrophysical Journal

\bibitem[{Martig {et~al.}(2016)Martig, Minchev, Ness, Fouesneau, \& Rix}]{martig2016}
Martig, M., Minchev, I., Ness, M., Fouesneau, M., \& Rix, H.-W. 2016, The Astrophysical Journal, 831, 139, \dodoi{10.3847/0004-637X/831/2/139}

\bibitem[{Martin {et~al.}(2014)Martin, Ibata, Rich, Collins, Fardal, Irwin, Lewis, McConnachie, Babul, Bate, Chapman, Conn, Crnojevi{\'c}, Ferguson, Mackey, Navarro, Pe{\~n}arrubia, Tanvir, \& {Valls-Gabaud}}]{martin2014}
Martin, N.~F., Ibata, R.~A., Rich, R.~M., {et~al.} 2014, The Astrophysical Journal, 787, 19, \dodoi{10.1088/0004-637X/787/1/19}

\bibitem[{McConnachie {et~al.}(2009)McConnachie, Irwin, Ibata, Dubinski, Widrow, Martin, C{\^o}t{\'e}, Dotter, Navarro, Ferguson, Puzia, Lewis, Babul, Barmby, Bienaym{\'e}, Chapman, Cockcroft, Collins, Fardal, Harris, Huxor, Mackey, Pe{\~n}arrubia, Rich, Richer, Siebert, Tanvir, {Valls-Gabaud}, \& Venn}]{mcconnachie2009}
McConnachie, A.~W., Irwin, M.~J., Ibata, R.~A., {et~al.} 2009, Nature, 461, 66, \dodoi{10.1038/nature08327}

\bibitem[{McConnachie {et~al.}(2018)McConnachie, Ibata, Martin, Ferguson, Collins, Gwyn, Irwin, Lewis, Mackey, Davidge, Arias, Conn, C{\^o}t{\'e}, Crnojevic, Huxor, Penarrubia, Spengler, Tanvir, {Valls-Gabaud}, Babul, Barmby, Bate, Bernard, Chapman, Dotter, Harris, McMonigal, Navarro, Puzia, Rich, Thomas, \& Widrow}]{mcconnachie2018}
McConnachie, A.~W., Ibata, R., Martin, N., {et~al.} 2018, The Astrophysical Journal, 868, 55, \dodoi{10.3847/1538-4357/aae8e7}

\bibitem[{McMonigal {et~al.}(2016)McMonigal, Lewis, Brewer, Irwin, Martin, McConnachie, Ibata, Ferguson, Mackey, \& Chapman}]{mcmonigal2016}
McMonigal, B., Lewis, G.~F., Brewer, B.~J., {et~al.} 2016, Monthly Notices of the Royal Astronomical Society, 461, 4374, \dodoi{10.1093/mnras/stw1657}

\bibitem[{Ogami {et~al.}(2024)Ogami, Tanaka, Komiyama, Chiba, Guhathakurta, Kirby, Wyse, Filion, Gilbert, Escala, Mori, Kirihara, Tanaka, Ishigaki, Hayashi, Lee, Sharma, Kalirai, \& Lupton}]{ogami2024a}
Ogami, I., Tanaka, M., Komiyama, Y., {et~al.} 2024, The Structure of the Stellar Halo of the {{Andromeda}} Galaxy Explored with the {{NB515}} for {{Subaru}}/{{HSC}}. {{I}}.: {{New Insights}} on the Stellar Halo up to 120 Kpc,  arXiv, \dodoi{10.48550/arXiv.2401.00668}

\bibitem[{Okamoto {et~al.}(2015)Okamoto, Arimoto, Ferguson, Bernard, Irwin, Yamada, \& Utsumi}]{okamoto2015}
Okamoto, S., Arimoto, N., Ferguson, A. M.~N., {et~al.} 2015, The Astrophysical Journal, 809, L1, \dodoi{10.1088/2041-8205/809/1/L1}

\bibitem[{Patel {et~al.}(2017)Patel, Besla, \& Mandel}]{patel2017}
Patel, E., Besla, G., \& Mandel, K. 2017, Monthly Notices of the Royal Astronomical Society, 468, 3428, \dodoi{10.1093/mnras/stx698}

\bibitem[{Pillepich {et~al.}(2014)Pillepich, Vogelsberger, Deason, {Rodriguez-Gomez}, Genel, Nelson, Torrey, Sales, Marinacci, Springel, Sijacki, \& Hernquist}]{pillepich2014a}
Pillepich, A., Vogelsberger, M., Deason, A., {et~al.} 2014, Monthly Notices of the Royal Astronomical Society, 444, 237, \dodoi{10.1093/mnras/stu1408}

\bibitem[{{Rodriguez-Gomez} {et~al.}(2016){Rodriguez-Gomez}, Pillepich, Sales, Genel, Vogelsberger, Zhu, Wellons, Nelson, Torrey, Springel, Ma, \& Hernquist}]{rodriguez-gomez2016}
{Rodriguez-Gomez}, V., Pillepich, A., Sales, L.~V., {et~al.} 2016, Monthly Notices of the Royal Astronomical Society, 458, 2371, \dodoi{10.1093/mnras/stw456}

\bibitem[{Schlafly {et~al.}(2012)Schlafly, Finkbeiner, Juri{\'c}, Magnier, Burgett, Chambers, Grav, Hodapp, Kaiser, Kudritzki, Martin, Morgan, Price, Rix, Stubbs, Tonry, \& Wainscoat}]{schlafly2012}
Schlafly, E.~F., Finkbeiner, D.~P., Juri{\'c}, M., {et~al.} 2012, The Astrophysical Journal, 756, 158, \dodoi{10.1088/0004-637X/756/2/158}

\bibitem[{Schlegel {et~al.}(1998)Schlegel, Finkbeiner, \& Davis}]{schlegel1998}
Schlegel, D.~J., Finkbeiner, D.~P., \& Davis, M. 1998, The Astrophysical Journal, 500, 525, \dodoi{10.1086/305772}

\bibitem[{Smercina {et~al.}(2023)Smercina, Dalcanton, Williams, Durbin, Lazzarini, Bell, Choi, Dolphin, Gilbert, Guhathakurta, Koch, Quirk, Rix, Rosolowsky, Seth, Skillman, \& Weisz}]{smercina2023}
Smercina, A., Dalcanton, J.~J., Williams, B.~F., {et~al.} 2023, The Astrophysical Journal, 957, 3, \dodoi{10.3847/1538-4357/acf3e8}

\bibitem[{Suzuki {et~al.}(2024)Suzuki, Chiba, Komiyama, Hayashi, Tanaka, Fukushima, Carlsten, Tokiwa, Qiu, \& Takada}]{suzuki2024}
Suzuki, Y., Chiba, M., Komiyama, Y., {et~al.} 2024, Publications of the Astronomical Society of Japan, \dodoi{10.1093/pasj/psae003}

\bibitem[{Takada {et~al.}(2014)Takada, Ellis, Chiba, Greene, Aihara, Arimoto, Bundy, Cohen, Dor{\'e}, Graves, Gunn, Heckman, Hirata, Ho, Kneib, F{\`e}vre, Lin, More, Murayama, Nagao, Ouchi, Seiffert, Silverman, Sodr{\'e}, Spergel, Strauss, Sugai, Suto, Takami, \& Wyse}]{takada2014}
Takada, M., Ellis, R.~S., Chiba, M., {et~al.} 2014, Publications of the Astronomical Society of Japan, 66, R1, \dodoi{10.1093/pasj/pst019}

\bibitem[{Tanaka {et~al.}(2017)Tanaka, Chiba, \& Komiyama}]{tanaka2017}
Tanaka, M., Chiba, M., \& Komiyama, Y. 2017, The Astrophysical Journal, 842, 127, \dodoi{10.3847/1538-4357/aa6d11}

\bibitem[{{The Astropy Collaboration} {et~al.}(2013){The Astropy Collaboration}, Robitaille, Tollerud, Greenfield, Droettboom, Bray, Aldcroft, Davis, Ginsburg, {Price-Whelan}, Kerzendorf, Conley, Crighton, Barbary, Muna, Ferguson, Grollier, Parikh, Nair, G{\"u}nther, Deil, Woillez, Conseil, Kramer, Turner, Singer, Fox, Weaver, Zabalza, Edwards, Azalee~Bostroem, Burke, Casey, Crawford, Dencheva, Ely, Jenness, Labrie, Lim, Pierfederici, Pontzen, Ptak, Refsdal, Servillat, \& Streicher}]{theastropycollaboration2013}
{The Astropy Collaboration}, Robitaille, T.~P., Tollerud, E.~J., {et~al.} 2013, Astronomy \& Astrophysics, 558, A33, \dodoi{10.1051/0004-6361/201322068}

\bibitem[{Thomas {et~al.}(2018)Thomas, McConnachie, Ibata, C{\^o}t{\'e}, Martin, Starkenburg, Carlberg, Chapman, Fabbro, Famaey, Fantin, Gwyn, {H{\'e}nault-Brunet}, Malhan, Navarro, Robin, \& Scott}]{thomas2018}
Thomas, G.~F., McConnachie, A.~W., Ibata, R.~A., {et~al.} 2018, Monthly Notices of the Royal Astronomical Society, 481, 5223, \dodoi{10.1093/mnras/sty2604}

\bibitem[{Tonry {et~al.}(2012)Tonry, Stubbs, Lykke, Doherty, Shivvers, Burgett, Chambers, Hodapp, Kaiser, Kudritzki, Magnier, Morgan, Price, \& Wainscoat}]{tonry2012}
Tonry, J.~L., Stubbs, C.~W., Lykke, K.~R., {et~al.} 2012, The Astrophysical Journal, 750, 99, \dodoi{10.1088/0004-637X/750/2/99}

\bibitem[{Van Der~Marel {et~al.}(2012)Van Der~Marel, Besla, Cox, Sohn, \& Anderson}]{vandermarel2012}
Van Der~Marel, R.~P., Besla, G., Cox, T.~J., Sohn, S.~T., \& Anderson, J. 2012, The Astrophysical Journal, 753, 9, \dodoi{10.1088/0004-637X/753/1/9}

\bibitem[{Van Der~Walt {et~al.}(2011)Van Der~Walt, Colbert, \& Varoquaux}]{vanderwalt2011}
Van Der~Walt, S., Colbert, S.~C., \& Varoquaux, G. 2011, Computing in Science \& Engineering, 13, 22, \dodoi{10.1109/MCSE.2011.37}

\bibitem[{Williams {et~al.}(2021)Williams, Durbin, Dalcanton, Lang, Girardi, Smercina, Dolphin, Weisz, Choi, Bell, Rosolowsky, Skillman, Koch, Lindberg, Hagen, Gordon, Seth, Gilbert, Guhathakurta, Lauer, \& Bianchi}]{williams2021}
Williams, B.~F., Durbin, M.~J., Dalcanton, J.~J., {et~al.} 2021, The Astrophysical Journal Supplement Series, 253, 53, \dodoi{10.3847/1538-4365/abdf4e}

\bibitem[{York {et~al.}(2000)York, Adelman, Anderson, Anderson, Annis, Bahcall, Bakken, Barkhouser, Bastian, Berman, Boroski, Bracker, Briegel, Briggs, Brinkmann, Brunner, Burles, Carey, Carr, Castander, Chen, Colestock, Connolly, Crocker, Csabai, Czarapata, Davis, Doi, Dombeck, Eisenstein, Ellman, Elms, Evans, Fan, Federwitz, Fiscelli, Friedman, Frieman, Fukugita, Gillespie, Gunn, Gurbani, {de Haas}, Haldeman, Harris, Hayes, Heckman, Hennessy, Hindsley, Holm, Holmgren, Huang, Hull, Husby, Ichikawa, Ichikawa, Ivezi{\'c}, Kent, Kim, Kinney, Klaene, Kleinman, Kleinman, Knapp, Korienek, Kron, Kunszt, Lamb, Lee, Leger, Limmongkol, Lindenmeyer, Long, Loomis, Loveday, Lucinio, Lupton, MacKinnon, Mannery, Mantsch, Margon, McGehee, McKay, Meiksin, Merelli, Monet, Munn, Narayanan, Nash, Neilsen, Neswold, Newberg, Nichol, Nicinski, Nonino, Okada, Okamura, Ostriker, Owen, Pauls, Peoples, Peterson, Petravick, Pier, Pope, Pordes, Prosapio, Rechenmacher, Quinn, Richards, Richmond, Rivetta, Rockosi, Ruthmansdorfer,
  Sandford, Schlegel, Schneider, Sekiguchi, Sergey, Shimasaku, Siegmund, Smee, Smith, Snedden, Stone, Stoughton, Strauss, Stubbs, SubbaRao, Szalay, Szapudi, Szokoly, Thakar, Tremonti, Tucker, Uomoto, Vanden~Berk, Vogeley, Waddell, Wang, Watanabe, Weinberg, Yanny, \& Yasuda}]{york2000}
York, D.~G., Adelman, J., Anderson, Jr., J.~E., {et~al.} 2000, The Astronomical Journal, 120, 1579, \dodoi{10.1086/301513}

\end{thebibliography}
\bibliographystyle{aasjournal}


\end{document}